\documentclass{article}

\usepackage{PRIMEarxiv}

\usepackage[utf8]{inputenc} % allow utf-8 input
\usepackage[T1]{fontenc}    % use 8-bit T1 fonts
\usepackage{hyperref}       % hyperlinks
\usepackage{url}            % simple URL typesetting
\usepackage{booktabs}       % professional-quality tables
\usepackage{amsfonts}       % blackboard math symbols
\usepackage{nicefrac}       % compact symbols for 1/2, etc.
\usepackage{microtype}      % microtypography
\usepackage{fancyhdr}       % header
\usepackage{ulem}           % for \sout (strikeout)
\usepackage{graphicx}       % graphics
\usepackage{natbib}
\usepackage{amsmath}
\usepackage{placeins}
\usepackage{listings}
\usepackage{xcolor}

\usepackage{tabularx}
\definecolor{codegreen}{rgb}{0,0.6,0}
\definecolor{codegray}{rgb}{0.4,0.4,0.4}
\definecolor{codepurple}{rgb}{0.58,0,0.82}
\definecolor{backcolour}{rgb}{0.98,0.98,0.97}
\definecolor{codecomment}{rgb}{0.4,0.4,0.4}
\lstdefinestyle{mystyle}{
	backgroundcolor=\color{backcolour},   
	commentstyle=\color{codecomment},
	keywordstyle=\color{codepurple},
	numberstyle=\tiny\color{gray},
	stringstyle=\color{codegreen},
	breakatwhitespace=false, 
	escapeinside={\%}{)},     
	breaklines=true,                 
	captionpos=b,                    
	keepspaces=true,                 
	numbers=left,                    
	numbersep=5pt,                  
	showspaces=false,                
	showstringspaces=false,
	showtabs=false,
	tabsize=2
}
\graphicspath{{media/}}     % organize your images and other figures under media/ folder

\graphicspath{{media/}}     % organize your images and other figures under media/ folder

\title{\textit{incompressibleFoam}: a new time consistent framework with BDF and DIRK integration schemes}

\author{
  Paulin FERRO*, \textcolor{black}{Pierre-Etienne MEILLER}, Paul LANDEL, Carla LANDRODIE, Marc PESCHEUX  \\
  SARL G-MET Ingénierie \\
  63 rue d'Hyères, 83140 Six-Fours-Les-Plages\\
  *corresponding author, paulin.ferro@g-met.fr \\
}

\begin{document}

\maketitle
\begin{abstract}
This work focuses on the development of a new incompressible solver, \textcolor{black}{\textit{incompressibleFoam}}, within OpenFOAM and integrating several numerical methods \textcolor{black}{within the same framework}. Two momentum interpolation (\textcolor{black}{NCMI/CMI}) methods are implemented, as well as two forms of the pressure Poisson equation \textcolor{black}{(corrected/standard)}. Regarding time discretization, \textcolor{black}{steady-state}, Backward Differentiation Formula (BDF) and the Singly Diagonally Implicit Runge-Kutta (SDIRK) methods, up to the third order, are implemented. The solver is tested on three benchmark cases to assess the performance of different numerical configurations. The results are also compared with the standard incompressible solver of OpenFOAM: \textit{pimpleFoam}. They provide perspective on previous attempts to improve OpenFOAM's incompressible solvers and give practical results regarding the choice of momentum interpolation methods, pressure equation formulations, and time discretization schemes. \textcolor{black}{It is found that the pressure-corrected form should be avoided while using the NCMI and that third order schemes are not superior to 2nd order schemes. The CMI should be privileged to avoid time step and relaxation factor dependence and pressure velocity decoupling.} Finally, the source code is released \textcolor{black}{in} the following github repository: https://github.com/ferrop/incompressibleFoam
\end{abstract}

% keywords can be removed
%\keywords{incompressible \and DIRK \and momentum interpolation \and OpenFOAM}


\section{Introduction}

High-fidelity CFD simulations involve the use of high-resolution space and time discretization methods. OpenFOAM (\cite{Weller1998}) offers an efficient open-source code at the interface between academic and industrial applications. OpenFOAM provides an efficient framework for implementing and testing various numerical algorithms on arbitrary polyhedral cell meshes. In the context of incompressible flows, recent efforts \textcolor{black}{(\cite{AGUERRE2020104396}, \cite{Komen2021}, \cite{ZHAO2024106458},\cite{GEORGE2024106426})} have been made to improve the accuracy, robustness, or consistency of standard transient incompressible OpenFOAM solvers: \textit{pimpleFoam} or \textit{pisoFoam} (based on the PISO loop \cite{Issa1986}). \textit{pimpleFoam} is the iterative version of \textit{pisoFoam} where momentum flux, turbulence, and explicit source terms are iterated over outer correctors.

\cite{Vuorinen2014} have developed \textit{rk4projectionFoam}, a solver that uses explicit fourth-order Runge-Kutta (ERK) with the Chorin fractional step method, \cite{chorin1968}. The solver has been tested for DNS channel flows with improved computational speed compared to  \textit{pisoFoam}. \textit{rk4projectionFoam} has also been successfully used for a Large-Eddy Simulation (LES) over the Bolund hill in Denmark, \cite{Vuorinen2015}. By investigating the source code, authors have noticed that after each Runge-Kutta stage, the flux used in the momentum equation is calculated by interpolating the velocity from cell center to cell face. This choice is not fully rigorous, as the resulting flux does not strictly satisfy the continuity equation. Regarding the Rhie \& Chow interpolation \cite{Rhie1983} \textcolor{black}{also called the momentum interpolation}, the old time face velocity is calculated by linear interpolation of the cell centered velocities. This approach is known to be non-consistent and lead to time step dependent solutions and pressure-velocity decoupling for small time steps \cite{Shen2001}.

\cite{KazemiKamyab2015} have implemented Explicit first-stage Singly Diagonally Implicit Runge–Kutta schemes (ESDIRK) with the PIMPLE loop in OpenFOAM. Their results indicate that the interpolation of the residual vector must be done carefully to preserve the temporal accuracy. Their algorithm uses conservative flux and a consistent Rhie \& Chow interpolation similar to the one proposed by \cite{BoYu2002}. 

A second-order non-iterative PISO solver has been proposed in \cite{tukovic2018consistent} with an extension to moving grids. In this solver, the consistent approach of \cite{Cubero2007} is used for the momentum interpolation and a 2nd order extrapolation is employed for the flux, velocity and pressure to maintain the 2nd order convergence. \cite{Lee2017} also showed that the 2nd order extrapolation improve the temporal accuracy with PISO loop. 

In the context of fluid structure interaction, \cite{Gillebaart2016} implemented BDF integration schemes up to order 3 with consistent momentum interpolation \cite{BoYu2002} and an under-relaxation strategy. 

Another attempt to improve the incompressible OpenFOAM solvers has been initiated by \cite{DAlessandro2018}. In their work, ERK and Singly Diagonally Implicit Runge–Kutta (SDIRK) with PISO loop and extension to heat transfer flows have been implemented. The third order DIRK solver \cite{dirk3Foam} is released. Inspection of the source code also shows the same arguable approach as in \cite{Vuorinen2014} with non-conservative momentum flux and non-consistent Rhie \& Chow interpolation.

In \cite{Komen2020} different momentum interpolation methods are compared for DIRK and ERK schemes in term of energy dissipation for the Taylor-Green vortex case. It has been found that the consistent \cite{BoYu2002} or \cite{Cubero2007} and OpenFOAM momentum interpolations are the most dissipative approaches. \textcolor{black}{These results are} in agreement with the observation of \cite{Bartholomew2018}. In contrast, the original approach of Rhie \& Chow interpolation, \cite{Rhie1983} better preserves the energy (but leads to time and relaxation factor dependent solutions). \cite{Komen2020} have also shown that a consistent momentum interpolation is the only method that preserves the formal temporal accuracy order, regardless of the mesh size.

An interesting approach is later developed in \cite{Komen2021} with a new PIMPLE based solver \textit{RKSymFoam} where symmetric discretization schemes are hard coded  (\textit{midPoint} interpolation, \textit{uncorrected} Laplacian and surface normal gradient). Two forms of the pressure equation are studied (referred as Chorin, \cite{chorin1968} and Van Kan, \cite{Kan1986} forms). Their work have shown that the Van Kan form is practically dissipation free and preserves second-order temporal convergence. \textcolor{black}{However, in \textit{RKSymFoam}, relaxations are not implemented and the PIMPLE algorithm is also hard coded without the tolerance error convergence criteria option. Moreover, the use of \textit{midPoint} interpolation schemes makes the solver likely unusable for industrial applications and complex meshes.}

\cite{Kim2022} have extended the SDIRK method for incompressible two-phase flows using the OpenFOAM approach for momentum interpolation. \textcolor{black}{Finally, \cite{TUKOVIC201270} have also implemented the consistent momentum interpolation of \cite{BoYu2002} in the context of interface tracking method}.

The developments of the solvers mentioned previously \textcolor{black}{single phase incompressible} are summarized in Table \ref{tab:allSolver}. To conclude, based on the available literature, a \textcolor{black}{robust} incompressible solver, providing a general framework for BDF and (E)SDIRK integration methods, with a consistent Rhie \& Chow interpolation, avoiding relaxation and time-step dependencies and with an open-source code is missing. In this context, \textcolor{black}{the main contributions of the paper are summarized below:
\begin{itemize}
	\item The development of a new open-source OpenFOAM solver \textit{incompressibleFoam} with two momentum interpolations (MI) and two pressure formulations based on literature. The solver should provide an efficient framework unifying BDF, (E)SDIRK and steady-state schemes. The main interest of grouping numerical methods within an unified framework is to reduce the maintenance efforts.
	\item The presented literature shows that recent efforts have been deployed to implement third and higher order schemes such as BDF3 or (E)SDIRK. However, it remains unclear if such schemes are relevant and bring significant advantages over classical 2nd order time schemes. Concerning the MI and/or the pressure form in the Poisson equation, previous studies are interested in some specific numerical aspects such as: dissipation, convergence order, robustness (occurrence of the checkerboard effect) or accuracy. But all of these criteria are not simultaneously covered. Hence, test cases are deployed to clarify these topics and to bring practical results for CFD developers as well as engineering applications. Finally, test case results and their comparison to reference data must allow to verify the code correctness of \textit{incompressibleFoam} as part of a validation exercise.
\end{itemize}
    }
%In the later text we refer to PISO as the algorithm proposed by \cite{Issa1986} and PIMPLE the iterative %version of PISO with relaxation handling (a merge between PISO and SIMPLE algorithms, \cite{Patankar1972}).

\begin{table}[!htbp]
	\centering
     \tiny
	\begin{tabularx}{\textwidth}{XXXXXXXX}
		
		\toprule
		Reference Context & Time integration method & Momentum Flux & Momentum interpolation & PV coupling & Relaxation & Moving mesh & Source code \\
		\midrule
		OpenFOAM \textit{pimpleFoam} / \textit{pisoFoam} \cite{Weller1998} & BDF1, BDF2, CrankNicolson & conservative & non-consistent & PISO / PIMPLE & yes & yes & within OpenFOAM   \\
    	\midrule
		Low dissipative. DNS, LES. \textit{rk4projectionFoam}, \cite{Vuorinen2014} & ERK & non-conservative & non-consistent & Chorin fractional step & no & no & see \cite{Vuorinen2015}   \\
		\midrule
		Consistent incompressible NS, \cite{KazemiKamyab2015} & ESDIRK & conservative & consistent, \cite{BoYu2002} & PIMPLE  & no & no & not available \\
		\midrule
		Consistent FSI, \cite{Gillebaart2016} & BDF1, BDF2, BDF3 & conservative & consistent, \cite{BoYu2002} & PIMPLE  & yes & no & not available \\  
		\midrule      		
		Consistent incompressible NS on moving grids, \cite{tukovic2018consistent} & BDF1, BDF2 & conservative & consistent, \cite{Cubero2007} & PISO  & no & yes & not available \\      
		\midrule  		
		ERK, and DIRK incompressible NS, \cite{DAlessandro2018} & ERK, SDIRK & non-conservative & non-consistent & PISO  & no & no & \cite{dirk3Foam} \\	    
		\midrule    		
		Symmetry-preserving PISO \textit{RKSymFoam}, \cite{Komen2020} & ERK, ESDIRK, SDIRK & conservative & non-consistent & PIMPLE  & no & no  & \cite{RKSymFoam} \\   
		\midrule
	    \textit{Consistent, incompressible, BDF, DIRK} & BDF1, BDF2, BDF3, ESDIRK, SDIRK & conservative & consistent, \cite{BoYu2002} & PIMPLE  & yes & no & \cite{incompressibleFoam} \\   		
		\bottomrule
	\end{tabularx}%
	\label{tab:allSolver}%
	\caption{incompressible algorithms implemented in OpenFOAM}
\end{table}%

\clearpage
\section{Acronyms and Nomenclature}
\begin{table}[h!]
\small
\centering
\color{black}
\begin{tabular}{ll}
\toprule
\textbf{Symbol} & \textbf{Description} \\
\midrule
\\
        BDF & Backward Differentiation Formula \\
        CFD  & Computational Fluid Dynamics \\
        CMI & Consistent Momentum Interpolation  \\
        DIRK & Diagonally Implicit Runge–Kutta  \\
        DNS & Direct Numerical Simulation \\
        ERK &  Explicit Runge-Kutta \\
        ESDIRK &  Explicit Singly Diagonally Implicit Runge–Kutta  \\
        LES & Large-Eddy Simulation \\
        MI & Momentum Interpolation\\
        NCMI & Non-Consistent Momentum Interpolation \\
        NS & Navier-Stokes \\
        PIMPLE  & Hybrid algorithm that combines PISO and SIMPLE \\
        PISO  & Pressure Implicit with Splitting of Operators \\
        SDIRK  & Singly Diagonally Implicit Runge-Kutta  \\
        SIMPLE  & Semi-Implicit Method for Pressure-Linked Equations \\
\\        
\bottomrule
\end{tabular}
\caption{Acronyms}
\end{table}

\begin{table}[h!]
\centering
\small
\begin{tabular}{ll}
\toprule
\textbf{Symbols} & \textbf{Descriptions}  \\
\midrule
\\
        \textbf{Latin letters}\\
        $a_{ij}$ & Coefficients of the Butcher matrix  \\
        $a_{N}$ & Momentum matrix off-diagonal coefficients   \\
        $a_P$ & Total coefficient $= a_s + a_t$   \\
        $a_s$ & Momentum matrix spatial diagonal coefficients  \\
        $a_t$ & Momentum matrix time diagonal coefficient   \\
        $c_i$ & Runge-Kutta abscissa: $c_i = \sum_j a_{ij}$   \\
        $d t$ & Time step  \\
        $d_{PN_f}$ & Distance between adjacent cell centers  \\
        $F$ & Explicit contribution of the temporal schemes \\
        $\textbf{H}$ &  Operator containing neighboring contributions and source terms  \\
        $\textbf{k}$ & Non-orthogonal contribution of the $\textbf{S}_f$ vector    \\
        $p$ & Kinematic pressure ($P/\rho$) \\
        $R_j(\textbf{u})$ & Residual at stage $j$  \\
        $\textbf{S}_f$ & Surface vector of face $f$  \\
        $\textbf{u}$ & Velocity field  \\
        $V_P$ & Volume of cell $P$ \\
\\
\textbf{Greek letters}\\
        $\alpha_U$ & Velocity under-relaxation factor   \\
        $\beta$ & Flux correction coefficient in OpenFOAM   \\
        \textbf{$\delta$} & Orthogonal contribution of the $\textbf{S}_f$ vector    \\
        $\epsilon_p$ & Pressure tolerance error   \\
        $\epsilon_U$ & Velocity tolerance error   \\
        $\nu$ & Kinematic viscosity  \\
        $\phi$ & Volumetric face flux \\
\\
        \textbf{Subscripts and additional notations}\\
        $n+1$ & Current time level index   \\
        $n-1$ & Previous time level   \\
        $k$ & Nonlinear iteration index   \\
        $i, j$ & DIRK or Runge-Kutta stage indices   \\
%        $*$ & Extrapolated (temporary) value   \\
\\
\bottomrule
\end{tabular}
\label{Nomenclature}
\caption{Nomenclature}
\end{table}

%\begin{table}[h!]
%    \centering
%    \begin{tabular}{ll}
%        \multicolumn{2}{l}{\textbf{Greek letters}} \\
%        \hline
%        \\
%        BDF & Backward Differentiation Formula \\
%       CFD  & Computational Fluid Dynamics \\
%        CMI & Consistent Momentum Interpolation  \\
%        DIRK & Diagonally Implicit Runge–Kutta  \\
%        DNS & Direct Numerical Simulation \\
%        ERK &  Explicit Runge-Kutta \\
%        ESDIRK &  Explicit Singly Diagonally Implicit Runge–Kutta  \\
%        MI & Momentum Interpolation\\
%        NCMI & Non-Consistent Momentum Interpolation \\
%        NS & Navier-Stokes \\
%        PIMPLE  & Hybrid algorithm that combines PISO and SIMPLE \\
%        PISO  & Pressure Implicit with Splitting of Operators \\
%        SDIRK  & Singly Diagonally Implicit Runge-Kutta  \\
%        SIMPLE  & Semi-Implicit Method for Pressure-Linked Equations \\
%        \\
%        \hline
%    \end{tabular}
%\end{table}

%%%%%%%%%%%%%%%%%%%%%%%%%%%%%%%%%%%%%%%%%%%%%%%%%%%%%%%%%%%%%%%%%%%%%%%%%%%%%%%%%%%%%%%%%%%%%%
%%%%%%%%%%%%%%%%%%%%%%%%%%%%%%%%%%%%%%%%%%%%%%%%%%%%%%%%%%%%%%%%%%%%%%%%%%%%%%%%%%%%%%%%%%%%%%
%%%%%%%%%%%%%%%%%%%%%%%%%%%%%%%%%%%%%%%%%%%%%%%%%%%%%%%%%%%%%%%%%%%%%%%%%%%%%%%%%%%%%%%%%%%%%%
\clearpage
\section{Mathematical and numerical procedure}

\subsection{Discretization of governing equations}
The Navier–Stokes equations for incompressible flows are given by the momentum and continuity equations:

	\begin{equation}
	\frac{\partial\boldsymbol{u}}{\partial t}+\boldsymbol{\nabla}\cdot(\boldsymbol{u}  \otimes \boldsymbol{u})=
	-\boldsymbol{\nabla}p+\boldsymbol{\nabla} \cdot (\nu \boldsymbol{\nabla} \boldsymbol{u}) + \boldsymbol{\nabla}\cdot\left[\nu dev2(\boldsymbol{\nabla}\boldsymbol{u}^{T})   \right]
    \end{equation}
	\begin{equation}
	\boldsymbol{\nabla}\cdot\boldsymbol{u}= 0
    \end{equation}

Where $\boldsymbol{u}$ is the velocity field, $p$ the kinematic pressure ($=P/\rho$) and $dev2$ is defined by \textcolor{black}{$dev2(\boldsymbol{\nabla}\boldsymbol{u}^{T}) = (\nabla\boldsymbol{u})^T - \frac{2}{3} \mathrm{tr}(\nabla  \boldsymbol{u})^T$)}. In the context of iterative PISO algorithm, all terms except the pressure gradient are discretized. Two subscripts are used: $n$ the time step and $k$ for a non-linear iteration within the same time step. For a given cell $N_P$, surrounded by neighboring cells $N_f$ sharing face $f$, the discretization reads, using Gauss theorem for convection and Laplacian operators:

  	\begin{equation}
  	  \left[\frac{\partial\boldsymbol{u}}{\partial t}\right]_{FVM} 
  	+ \underset{f}{\sum} \phi^k \boldsymbol{u}_f^{n+1} 
  	=
  	 \underset{f}{\sum} \nu \boldsymbol{\nabla}\boldsymbol{u}_f^{n+1} \cdot \boldsymbol{S}_f
  	+\underset{f}{\sum} \nu \left[dev2((\boldsymbol{\nabla}\boldsymbol{u}^k)^{T}) \right]_{f}
  	-\textcolor{black}{\left\{\boldsymbol{\nabla}p^{n+1}\right\}}
    \end{equation}

Where the bracket $[.]_{FVM}$ indicates a discretization scheme that will be described in sections \ref{BDF} and \ref{DIRK} \textcolor{black}{and $\{.\}$ a symbolic notation indicating that the pressure gradient is left un-discretized}. $\phi^k$ is the volumetric face flux which is updated within the non-linear iteration. The bracket $[.]_{f}$ indicates a cell center to face center linear interpolation. For the convective term, depending on the chosen discretization scheme, $\boldsymbol{u}_f$ is written as a function of cell velocities $\boldsymbol{u}_P$ and $\boldsymbol{u}_{N_{f}}$, feeding the matrix diagonal and off-diagonal coefficients. The mesh non-orthogonality is handled using the over-relaxed approach, \citet{Jasak1996}. The surface vector $\boldsymbol{S}_{f}$ is then decomposed into two parts, the orthogonal $\boldsymbol{\delta}$ and non-orthogonal $\boldsymbol{k}$ contributions:

	\begin{equation}
	\boldsymbol{\nabla}\boldsymbol{u}_f^{n+1}\cdot\boldsymbol{S}_{f}
	=
	 \underset{implicit}{\underbrace{\boldsymbol{\nabla}\boldsymbol{u}_f\cdot\boldsymbol{\delta}}}
	+\underset{explicit}{\underbrace{\boldsymbol{\nabla}\boldsymbol{u}_f\cdot\boldsymbol{k}}}
	=
	 \frac{\boldsymbol{u}^{n+1}_P-\boldsymbol{u}^{n+1}_{N_f}}{d_{PN_{f}}} ||\boldsymbol{\delta}||
	 +[\boldsymbol{\nabla}\boldsymbol{u}^{k-1}\cdot\boldsymbol{k} ]_{f}	\label{eq:laplacianDiscr}
 	\end{equation}

The orthogonal part is discretized implicitly, contributing to matrix diagonal and off-diagonal coefficients, while the non-orthogonal contribution is treated explicitly and added to the matrix source term. After the discretization process, the momentum equation can be rewritten in an algebraic form.

\begin{equation}
	  \left[\frac{\partial\boldsymbol{u}}{\partial t}\right]_{FVM}
	=- a_{s}\boldsymbol{u}^{n+1}_{P}
	-\underset{f}{\sum}a_{N_{f}}\boldsymbol{u}^{n+1}_{N_{f}}
	+ \boldsymbol{s}(\boldsymbol{u})
	- \textcolor{black}{\left\{\boldsymbol{\nabla}p^{n+1}\right\}}
	=- a_{s}\boldsymbol{u}^{n+1}_{P}
	+ \boldsymbol{H}(\boldsymbol{u})
	- \textcolor{black}{\left\{\boldsymbol{\nabla}p^{n+1}\right\}} \label{eq:momentum}
\end{equation}

Where $a_s$ is the diagonal coefficient arising from the spatial schemes, $a_{N}$ the off-diagonal coefficients, and $\boldsymbol{s}(\boldsymbol{u})$ groups all the explicit terms (non-orthogonal correction, deferred correction etc...). The $\boldsymbol{H}$ operator is introduced and contains all the source terms except the old-time contributions. The optional resolution of Equation \ref{eq:momentum} is the \textit{momentumPredictor} step in OpenFOAM. In this work, the \textit{momentumPredictor} step is always solved.
%%%%%%%%%%%%%%%%%%%%%%%%%%%%%%%%%%%%%%%%%%%%%%%%%%%%%%%%%%%%%%%%%%%%%%%%%%%%%%%%%%%%%%%%%%%%%%
%%%%%%%%%%%%%%%%%%%%%%%%%%%%%%%%%%%%%%%%%%%%%%%%%%%%%%%%%%%%%%%%%%%%%%%%%%%%%%%%%%%%%%%%%%%%%%
%%%%%%%%%%%%%%%%%%%%%%%%%%%%%%%%%%%%%%%%%%%%%%%%%%%%%%%%%%%%%%%%%%%%%%%%%%%%%%%%%%%%%%%%%%%%%%
\subsection{Backward differentiation formula}\label{BDF}
Backward differencing formula (BDFi) schemes, from order i = 1 to 3, are considered for the time integration of NS equations. While BDF1 and BDF2 are A-stable, the BDF3 scheme is only conditionally stable for implicit time integration. The BDF schemes are written below with constant time stepping for the sake of clarity but implemented in \textit{incompressibleFoam} with variable time stepping.

  	\begin{equation}
	 \left[\frac{\partial\boldsymbol{u}}{\partial t}\right]_{FVM} = \frac{\boldsymbol{u}^{n+1}_P - \boldsymbol{u}^{n}_P }{dt}V_P 
    \end{equation}

  	\begin{equation}
	 \left[\frac{\partial\boldsymbol{u}}{\partial t}\right]_{FVM} = \frac{3\boldsymbol{u}^{n+1}_P - 4\boldsymbol{u}^{n}_P +\boldsymbol{u}^{n-1}_P  }{2dt}V_P 
    \end{equation}

  	\begin{equation}
	 \left[\frac{\partial\boldsymbol{u}}{\partial t}\right]_{FVM} = \frac{11\boldsymbol{u}^{n+1}_P - 18\boldsymbol{u}^{n}_P +9\boldsymbol{u}^{n-1}_P -2\boldsymbol{u}^{n-2}_P}{6d t} V_P
    \end{equation}

The momentum equation is obtained by combining Equation \ref{eq:momentum} and BDF schemes presented above. Then, under-relaxation is applied for increasing the diagonal dominance. 

  	\begin{equation}
	\boldsymbol{u}^{n+1}_P
	=	 
	\alpha_U\left(\frac{\boldsymbol{H}(\boldsymbol{u}^{n+1})
		- \textcolor{black}{\left\{\boldsymbol{\nabla}p^{n+1}\right\}} 
		+ F(\boldsymbol{u}_{P}^n,\boldsymbol{u}_{P}^{n-1},..)}{(a_t + a_s )} 
	   \right)
	+ \left(1-\alpha_U \right)\boldsymbol{u}^{k-1}_P \label{eq:momentumBDF}
    \end{equation}

Where $\alpha_U$ is the relaxation factor, $a_t$ the diagonal coefficient arising from the temporal scheme and $F$ a linear function of the old time cell centered velocities. 

%%%%%%%%%%%%%%%%%%%%%%%%%%%%%%%%%%%%%%%%%%%%%%%%%%%%%%%%%%%%%%%%%%%%%%%%%%%%%%%%%%%%%%%%%%%%%%
%%%%%%%%%%%%%%%%%%%%%%%%%%%%%%%%%%%%%%%%%%%%%%%%%%%%%%%%%%%%%%%%%%%%%%%%%%%%%%%%%%%%%%%%%%%%%%
%%%%%%%%%%%%%%%%%%%%%%%%%%%%%%%%%%%%%%%%%%%%%%%%%%%%%%%%%%%%%%%%%%%%%%%%%%%%%%%%%%%%%%%%%%%%%%

\subsection{Diagonally Implicit Runge–Kutta}\label{DIRK}
Diagonally Implicit Runge–Kutta (DIRK) are a class of methods with a lower triangular Butcher matrix. DIRK methods with the same coefficient on the diagonal are known as Singly Diagonally Rune Kutta (SDIRK). A DIRK integration is composed of $N$ stages, $N$ being the size of the Butcher matrix. Each stage, labeled with the counter $i$, is formerly a BDF1 integration with a source term. The same form is adopted as in \cite{Komen2020} where the pressure gradient is not included in the residual vector. This is possible because the pressure gradient acts as a Lagrangian multiplier, \cite{Sanderse2012}. Preliminary tests conducted in this work have shown that this approach is mandatory in term of reliability, avoiding pressure oscillations at the start of the simulations.

  	\begin{equation}
	\frac{\boldsymbol{u}^{i}_P - \boldsymbol{u}^{n}_P }{d t} V_P =
	a_{ii}\left( 
  	- a_{s}\boldsymbol{u}^{i}_{P}
  	+ \boldsymbol{H}(\boldsymbol{u}^{i})\right)
    + \underset{j<i}{\sum} a_{ij}\boldsymbol{R}_j(\boldsymbol{u}^{j}_P)
    - c_i \textcolor{black}{\left\{\boldsymbol{\nabla}p^{n+1}\right\}}  
    \end{equation}

Where $a_{ij}$ are the coefficients of the Butcher matrix and $c_i = \underset{j \le i}{\sum}a_{ij}$. The residual vector $\boldsymbol{R}_j(\boldsymbol{u}_P)$ is calculated at the end of a the stage $j$.

  	\begin{equation}
	\boldsymbol{R}_j(\boldsymbol{u}^{j}_P) =
	- a_{s}\boldsymbol{u}^{j}_{P}
    + \boldsymbol{H}(\boldsymbol{u}^{j})
	\label{eq:residuals}
    \end{equation}

Under-relaxation is applied as for BDF1 scheme:

  	\begin{equation}
	 \boldsymbol{u}^{i}_P
	 =	 
	\alpha_U\left( \frac{
		a_{ii}\boldsymbol{H}(\boldsymbol{u}^{i})
		- c_i \textcolor{black}{\left\{\boldsymbol{\nabla}p^{n+1}\right\}} 
		+ \underset{j<i}{\sum} a_{ij}\boldsymbol{R}_j(\boldsymbol{u}^{j}_P)
		+ \frac{V_P}{d t} \boldsymbol{u}^{n}_P }{(a_t + a_{ii}a_s) } \right)
	+ \left(1-\alpha_U \right) \boldsymbol{u}^{k-1}_P \label{eq:momentumRK}
    \end{equation}

In this work, \textcolor{black}{the} stiffly accurate DIRK method is used, hence the new field values are the ones calculated from the last stage:

\begin{equation}
	\boldsymbol{u}^{n+1}_P = \boldsymbol{u}^{N}_P
\end{equation}
\begin{equation}
	p^{n+1}_P = p^{N}_P
\end{equation}
\begin{equation}
	\phi_{f}^{n+1} = \phi_{f}^{N}
\end{equation}

%%%%%%%%%%%%%%%%%%%%%%%%%%%%%%%%%%%%%%%%%%%%%%%%%%%%%%%%%%%%%%%%%%%%%%%%%%%%%%%%%%%%%%%%%%%%%%
%%%%%%%%%%%%%%%%%%%%%%%%%%%%%%%%%%%%%%%%%%%%%%%%%%%%%%%%%%%%%%%%%%%%%%%%%%%%%%%%%%%%%%%%%%%%%%
%%%%%%%%%%%%%%%%%%%%%%%%%%%%%%%%%%%%%%%%%%%%%%%%%%%%%%%%%%%%%%%%%%%%%%%%%%%%%%%%%%%%%%%%%%%%%%

\subsection{Momentum interpolation}\label{momentumInterpolation}

The Poisson equation for the pressure is obtained by imposing the continuity constrain to Equations \ref{eq:momentumBDF} and \ref{eq:momentumRK}. To restrain checkerboard oscillations on collocated grids, the so-called \cite{Rhie1983} interpolation is used to obtain the face velocity by mimicking Equations \ref{eq:momentumBDF} and \ref{eq:momentumRK}. In order to avoid relaxation factor and time step dependencies, the interpolation needs to be done in a consistent way, \cite{BoYu2002}, \cite{Cubero2007}, \cite{Pascau2011}. Two approaches are followed in this work:
\begin{itemize}
	\item The first approach is the one proposed by \cite{BoYu2002} to avoid relaxation factor and time step dependencies. Instead of interpolating the previous and old times velocities from cell center to face center, the corresponding conservative face flux $\phi_f$ are used. This approach is referred as the \textit{consistent momentum interpolation}: \textbf{CMI}.
	\item The second approach has the same form as the \textit{consistent} one. However, the old time flux is calculated by direct linear interpolation of the cell centered old time velocities $\phi_f = [\boldsymbol{u}^{n}_{P}]_f\cdot\boldsymbol{S}_{f}$ to the face. This approach is referred latter as the \textit{non-consistent momentum interpolation}: \textbf{NCMI}.
\end{itemize}

 The interpolation is written below in term of face flux for both BDF and Runge-Kutta integration schemes. The $\boldsymbol{H}$ operator is calculated using the available velocity field $\boldsymbol{u}^{*}_{P}$ and thus, $\boldsymbol{H}$ is iterated within each corrector of the PISO loop.

For BDF integration:
\begin{equation}
	 \phi^{*}_{f}
	=\alpha_U \left(-\frac{\boldsymbol{\nabla}p_f^{n+1}}{[a_P]_f}\cdot\boldsymbol{S}_{f}
	 +\frac{[\boldsymbol{H}(\boldsymbol{u}^{*}_{P})]_f\cdot\boldsymbol{S}_{f}}{[a_P]_f}
	 +\frac{F(\phi_{f}^n,\phi_{f}^{n-1},..)}{[a_P]_f} \right) 
	 +(1 - \alpha_U)\phi_{f}^{k-1} \label{eq:faceMomentumBDF}
\end{equation}

Where $a_p = a_t + a_s$ and $[.]_{f}$ indicates linear interpolation from cell center to face center \textcolor{black}{and $\boldsymbol{\nabla}p_f$ is the pressure surface normal gradient}. For DIRK integration:
\begin{equation}
	\phi^{*}_{f}
	=\alpha_U \left(-\frac{c_i}{[a_P]_f}\boldsymbol{\nabla}p_f^{n+1}\cdot\boldsymbol{S}_{f}
	+\frac{a_{ii}[\boldsymbol{H}(\boldsymbol{u}^{*}_{P})]_f\cdot\boldsymbol{S}_{f}}{[a_P]_f} + \frac{\underset{j<i}{\sum} a_{ij}R_{f,j}(\phi_{f}^{j})}{[a_P]_f} 
	+\frac{V_P\phi_{f}^{n}}{d t[a_P]_f} \right)
	+(1 - \alpha_U)\phi_{f}^{k-1} \label{eq:faceMomentumDIRK}
\end{equation}

Where $a_P = \frac{V_P}{d t} + a_{ii}a_s $. The face residuals flux $R_{f,j}(\phi_{f}^{j})$ is calculated in a different way depending on the momentum interpolation approach. 

\begin{itemize}
	\item For the CMI approach, $R_{f,j}(\phi_{f}^{j})$ has to be calculated in a consistent way as suggested by \cite{KazemiKamyab2015} from Equation \ref{eq:residuals}:
	\begin{equation}
		R_{f,j}(\phi_{f}^{j}) =
		- [a_{s}]_f\phi_{f}^{j}
		+ [\boldsymbol{H}(\boldsymbol{u}^{j}_{P})]_f\cdot\boldsymbol{S}_{f}\label{eq:faceResiduals}
	\end{equation}
	\item For the NCMI approch, $R_{f,j}(\phi_{f}^{j})$ is calculated by linear interpolation of the cell centered residuals to the face mesh: 
         \begin{equation}
             R_{f,j}(\phi_{f}^{j}) = [\boldsymbol{R}_j(\boldsymbol{u}^{j}_P)]_f \cdot\boldsymbol{S}_{f} 
         \end{equation}
\end{itemize}

In the standard incompressible OpenFOAM solvers (i.e \textit{pimpleFoam}), the momentum interpolation is done otherwise. The spatial and temporal contributions ($a_t$ and $a_s$) are not split and thus, $\boldsymbol{H}$ operator contains old time and previous iteration fields. The interpolation of each term is performed according to the procedure of \cite{Rhie1983}. Then, a flux correction is added to replace the old time contribution by the corresponding conservative flux through the \textit{fvc::ddtCorr} function. This correction aims at reducing the decoupling between pressure and velocity. Hence, the OpenFOAM momentum interpolation has the following form (with first order Euler scheme):

\begin{equation}
\phi^{*}_{f_{OF}} =
-\left[\frac{1}{a_P} \right ]_f \boldsymbol{\nabla}p_f^{n+1}\cdot\boldsymbol{S}_{f} +\left[\frac{\boldsymbol{H}(\boldsymbol{u}^{*}_{P})}{a_P}\right]_f\cdot\boldsymbol{S}_{f}
+ \beta \left[\frac{1}{a_P} \right ]_f \frac{\left(\phi^{n}_{f} -  [\boldsymbol{u}^{n}_{P}]_f\cdot\boldsymbol{S}_{f}\right)}{d t}
 \label{eq:faceMomentumOF}
\end{equation}

Where $\beta$ is a scale factor calculated as follow (function \textit{fvcDdtPhiCoeff}):

\begin{equation}
	\beta = 1 - min\left(\frac{|\phi^{n}_{f} - [\boldsymbol{u}^{n}_{P}]_f\cdot\boldsymbol{S}_{f}|}{|\phi^{n}_{f}|},1\right)
\end{equation}
The activation of the flux correction can be controlled by the user with the keyword \textit{ddtCorr} (true/false - default value is true). It is notable that, since $\left[\frac{1}{a_P} \right ]_f [\boldsymbol{u}^{n}_{P}]_f\cdot\boldsymbol{S}_{f} \neq \left[\frac{ \boldsymbol{u}^{n}_{P}\cdot\boldsymbol{S}_{f}}{a_P} \right ]_f $ and irrespective to the value of $\beta$, the old contribution is not entirely canceled from the momentum interpolation and the OpenFOAM approach \textcolor{black}{will lead to time-step and relaxation factor dependent results}.

\textcolor{black}{The fundamental difference between CMI and NCMI is the error introduced by the interpolation procedure. For the sake of clarity a BDF integration without relaxation factor is assumed. The error can be calculated by comparing the face velocity from Equation \ref{eq:faceMomentumBDF} (written in term of velocity) :}
\textcolor{black}{
\begin{align}u_f = \left\{\begin{array}{ll}
      -\frac{\boldsymbol{\nabla}p_f^{n+1}}{[a_P]_f}
	 +\frac{[\boldsymbol{H}(\boldsymbol{u}^{*}_{P})]_f}{[a_P]_f}
	 +\frac{F(\left[\boldsymbol{u}^n_P\right]_f,\left[\boldsymbol{u}^{n-1}_P\right]_f,..)}{[a_P]_f} & \text{NCMI} \\
      -\frac{\boldsymbol{\nabla}p_f^{n+1}}{[a_P]_f}
	 +\frac{[\boldsymbol{H}(\boldsymbol{u}^{*}_{P})]_f}{[a_P]_f}
	 +\frac{F(\boldsymbol{u}^n_f,\boldsymbol{u}^{n-1}_f,..)}{[a_P]_f} & \text{CMI}
 \end{array}\right.\end{align}}

\textcolor{black}{
to an interpolation of Equation \ref{eq:momentumBDF} from cell center to face (the coefficients are interpolated in the same way as in Equation \ref{eq:faceMomentumBDF}) :
     \begin{align} \left[\boldsymbol{u}_P\right]_f = -\frac{\left[\boldsymbol{\nabla}p^{n+1}\right]_f}{[a_P]_f}
	 +\frac{[\boldsymbol{H}(\boldsymbol{u}^{*}_{P})]_f}{[a_P]_f}
	 +\frac{F(\left[\boldsymbol{u}^n_P\right]_f,\left[\boldsymbol{u}^{n-1}_P\right]_f,..)}{[a_P]_f} \end{align}
\textcolor{black}{     
For the NCMI the face velocity error is given by:}
\textcolor{black}{
\begin{align}
	 u_{f} - \left[\boldsymbol{u}_P\right]_f &=  \frac{1}{[a_P]_f}
\left(\left[\boldsymbol{\nabla}p^{n+1}\right]_f-\boldsymbol{\nabla}p_f^{n+1}\right)\propto d t d x^2 {\nabla}^3p = O(d t d x ^2) \label{eq:errorNCMI}
\end{align}
%The face velocity contain errors of the first order in time and second order in space.
}
\textcolor{black}{
And for the CMI:
\begin{equation}
	 \boldsymbol{u}_{f} - \left[\boldsymbol{u}_P\right]_f = 
      \frac{1}{[a_P]_f} \left(\left[\boldsymbol{\nabla}p^{n+1}\right]_f-\boldsymbol{\nabla}p_f^{n+1}\right) 
      + \frac{1}{[a_P]_f} \left(\frac{\boldsymbol{u}_f^n-\left[\boldsymbol{u}^n\right]_f}{d t}\right)
      \propto d t d x^2 {\nabla}^3p + d x^2 {\nabla}^2 u = O(d x ^2) \label{eq:errorCMI}
\end{equation}}
Hence,
\textcolor{black}{
\begin{itemize}
    \item The NCMI is equivalent to adding a term proportional to a third derivative of the pressure. This term acts as a filter avoiding the pressure oscillations (i.e. checkerboard). However, this term tends to 0 as the time step decreases, explaining why checkerboard can occur for small time steps. 
    \item For the CMI, there is another term that is proportional to a second derivative of the velocity. The CMI adds a time-step independent filter ensuring that checkerboard oscillations will not appear for small time steps. 
  \end{itemize}
An alternative form is to solve the corrected pressure $p_c$ where $p^{n+1} = p^{n} + p_c^{n+1}$ and $p_c = O(d t)$, \cite{vanDoormaal1984}, \cite{Kan1986}. In this case, we can show that the error becomes $O(d t^2 d x^2)$ for the NCMI and remains $O(d x^2)$ for the CMI. Hence, the pressure-corrected form is more sensitive to checkerboard oscillations unless a CMI is used. \cite{ham2004energy} have shown that the face velocity error is responsible of a sink term in the kinetic energy balance. The Table \ref{tab:interpError} summarizes the face velocity errors for each combination of MI and pressure form. Regardless of the pressure form we can expect the CMI formulation to be the most dissipative. In opposition the NCMI with corrected pressure should be the less dissipative.}
 }

\begin{table}[!htbp]
	\centering
     \tiny
	\begin{tabularx}{\textwidth}{>{\color{black}}X >{\color{black}}X >{\color{black}}X >{\color{black}}X >{\color{black}}X}
		\toprule
		& NCMI standard & NCMI pressure corrected & CMI standard & CMI pressure corrected \\
		\midrule
     Interpolation error  & $O(d t d x^2 )$ & $O(d t^2 d x^2) $ & $O(d x^2)$ & $O(d x^2)$ \\
 		\midrule
     Limit ($dt \rightarrow 0$)  & $d t $ & $d t^2$ & Constant & Constant \\
	\bottomrule
\end{tabularx}
    \caption{Interpolation errors for MI methods}
    \label{tab:interpError}    
\end{table}

%%%%%%%%%%%%%%%%%%%%%%%%%%%%%%%%%%%%%%%%%%%%%%%%%%%%%%%%%%%%%%%%%%%%%%%%%%%%%%%%%%%%%%%%%%%%%%
%%%%%%%%%%%%%%%%%%%%%%%%%%%%%%%%%%%%%%%%%%%%%%%%%%%%%%%%%%%%%%%%%%%%%%%%%%%%%%%%%%%%%%%%%%%%%%
%%%%%%%%%%%%%%%%%%%%%%%%%%%%%%%%%%%%%%%%%%%%%%%%%%%%%%%%%%%%%%%%%%%%%%%%%%%%%%%%%%%%%%%%%%%%%%

\subsection{Pressure Poisson equation}\label{poisson}
The Poisson equation for pressure is obtained by applying the continuity equation to Equations \ref{eq:faceMomentumBDF} and \ref{eq:faceMomentumDIRK}, giving for the BDF and DIRK integration schemes:
\begin{equation}
	\underset{f}{\sum} \frac{\alpha_U }{[a_P]_f}{\nabla}p_f^{n+1}\cdot\boldsymbol{S}_{f}
	=
	\underset{f}{\sum}\alpha_U\left(\frac{[\boldsymbol{H}(\boldsymbol{u}^{*}_{P})]_f\cdot\boldsymbol{S}_{f}}{[a_P]_f}
	+\frac{F(\phi_{f}^n,\phi_{f}^{n-1},..)}{[a_P]_f} \right)
	+\underset{f}{\sum}(1 - \alpha_U)\phi_{f}^{k-1} \label{eq:poissonBDF}
\end{equation}
\begin{equation}
	\underset{f}{\sum} c_i\frac{\alpha_U }{[a_P]_f}\boldsymbol{\nabla}p_f^{n+1}\cdot\boldsymbol{S}_{f}
	=
	\underset{f}{\sum}\alpha_U\left(\frac{a_{ii}[\boldsymbol{H}(\boldsymbol{u}^{*}_{P})]_f\cdot\boldsymbol{S}_{f}}{[a_P]_f} + \frac{\underset{j<i}{\sum} a_{ij}R_{f,j}(\phi_{f}^{j})}{[a_P]_f} 
	+\frac{V_P\phi_{f}^{n}}{d t[a_P]_f} \right)
	+\underset{f}{\sum}(1 - \alpha_U)\phi_{f}^{k-1} \label{eq:poissonDIRK}
\end{equation} 
Where the face pressure gradient $\boldsymbol{\nabla}p_f^{n+1}$ is discretized using Equation \ref{eq:laplacianDiscr} with implicit orthogonal and explicit non-orthogonal contributions. In this study, two different forms of the pressure equation are implemented :

\begin{itemize}
	\item Form 1: the pressure equation is solved as in Equations \ref{eq:poissonBDF} or \ref{eq:poissonDIRK}. This form is referred to as the \textit{\textbf{standard}} one.
	\item Form 2: the pressure is split as follows $p^{n+1} = p^{n} + p_c^{n+1}$ (\cite{vanDoormaal1984}, \cite{Kan1986}) where $p_c$ is the pressure correction being solved for the Poisson equation. This form is referred to the \textit{\textbf{corrected}} one. The old time pressure $p^{n}$ part generates a source term $\underset{j}{\sum} c_i\frac{\alpha_U}{[a_P]_f}\boldsymbol{\nabla}p_f^{n}\cdot\boldsymbol{S}_{f}$ that is treated differently regarding the momentum interpolation approach. For the CMI, the old time pressure gradient flux $\boldsymbol{\nabla}p_f^{n}\cdot\boldsymbol{S}_{f}$ is evaluated explicitly at the faces using equation \ref{eq:laplacianDiscr}. For the NCMI, the old time pressure gradient flux is calculated by a direct linear interpolation of the old time cell centered pressure gradient to the face. The pressure corrected form implies that the pressure boundary conditions need to be defined in term of $p_{c}$ instead of $p$. This can lead to a complex implementation as the user shall define $p$ instead of $p_c$ from a practical point of view. 
\end{itemize}

%%%%%%%%%%%%%%%%%%%%%%%%%%%%%%%%%%%%%%%%%%%%%%%%%%%%%%%%%%%%%%%%%%%%%%%%%%%%%%%%%%%%%%%%%%%%%%
%%%%%%%%%%%%%%%%%%%%%%%%%%%%%%%%%%%%%%%%%%%%%%%%%%%%%%%%%%%%%%%%%%%%%%%%%%%%%%%%%%%%%%%%%%%%%%
%%%%%%%%%%%%%%%%%%%%%%%%%%%%%%%%%%%%%%%%%%%%%%%%%%%%%%%%%%%%%%%%%%%%%%%%%%%%%%%%%%%%%%%%%%%%%%

\subsection{Velocity and flux update}

 After the resolution of the Poisson equation, velocity and flux are updated using the new pressure values. 
\begin{itemize}
	\item The velocity update is done trough Equations \ref{eq:momentumBDF} or \ref{eq:momentumRK} depending on the time integration method. The pressure gradient is calculated using an user defined scheme: cell centered Gauss or least squares schemes.
	\item The face flux is updated using Equations \ref{eq:faceMomentumBDF} or \ref{eq:faceMomentumDIRK} depending on the time integration method. It has to be noticed that the pressure gradient flux $\left[\boldsymbol{\nabla}p_f^{n+1}\right]\cdot\boldsymbol{S}_{f}$ (also present in Equation \ref{eq:faceResiduals}) is directly calculated using Poisson equation coefficients and the flux contribution from the non-orthogonal correction. This ensures that the updated flux verify the continuity equation in opposition to the method used in \cite{Vuorinen2014} or \cite{DAlessandro2018} where face flux are linearly interpolated $\phi_{f}^{n+1} = [\boldsymbol{u}^{n+1}]_f\cdot\boldsymbol{S}_{f}$.
\end{itemize}
%%%%%%%%%%%%%%%%%%%%%%%%%%%%%%%%%%%%%%%%%%%%%%%%%%%%%%%%%%%%%%%%%%%%%%%%%%%%%%%%%%%%%%%%%%%%%%
%%%%%%%%%%%%%%%%%%%%%%%%%%%%%%%%%%%%%%%%%%%%%%%%%%%%%%%%%%%%%%%%%%%%%%%%%%%%%%%%%%%%%%%%%%%%%%
%%%%%%%%%%%%%%%%%%%%%%%%%%%%%%%%%%%%%%%%%%%%%%%%%%%%%%%%%%%%%%%%%%%%%%%%%%%%%%%%%%%%%%%%%%%%%%
\subsection{Fields extrapolation}
\cite{tukovic2018consistent} and \cite{Lee2017} have shown that with PISO algorithm, a 2nd order extrapolation can improve the temporal \textcolor{black}{convergence order}. At the beginning of a time loop, a 2nd order extrapolation is implemented for flux, pressure and velocity with a variable time step approach. The linear extrapolation is extended to be consistent with DIRK integration where the stage time step is given by $c_i d t$.

\begin{equation}
\begin{split} 
	&\boldsymbol{u}^{i}_P = \frac{c_i d t}{d t_0}(\boldsymbol{u}^{n}_P-\boldsymbol{u}^{n-1}_P)+\boldsymbol{u}^{n}_P \\
     & p^{i}_P = \frac{c_i d t}{d t_0}(p^{n}_P-p^{n-1}_P)+p^{n}_P \\	
     & \phi^{i}_f = \frac{c_i d t}{d t_0}(\phi^{n}_f-\phi^{n-1}_f)+\phi^{n}_f    \label{eq:fieldExtrapolation}
\end{split}
\end{equation}

For BDF integration, $i=n+1$ and $c_i = 1$.

%%%%%%%%%%%%%%%%%%%%%%%%%%%%%%%%%%%%%%%%%%%%%%%%%%%%%%%%%%%%%%%%%%%%%%%%%%%%%%%%%%%%%%%%%%%%%%
%%%%%%%%%%%%%%%%%%%%%%%%%%%%%%%%%%%%%%%%%%%%%%%%%%%%%%%%%%%%%%%%%%%%%%%%%%%%%%%%%%%%%%%%%%%%%%
%%%%%%%%%%%%%%%%%%%%%%%%%%%%%%%%%%%%%%%%%%%%%%%%%%%%%%%%%%%%%%%%%%%%%%%%%%%%%%%%%%%%%%%%%%%%%%

\subsection{\textit{incompressibleFoam} - algorithm}
 The coexistence between BDF and DIRK integration in the same solver is ensured through the dimension of the butcher table. For BDF schemes, the dimension of butcher table is equal to one, hence only one Runge-Kutta loop is performed. For each Runge-Kutta stage, the solver takes advantage of the PIMPLE algorithm. At the beginning of the time step, the fields are extrapolated (optionally) using equation \ref{eq:fieldExtrapolation}. As mentioned above, this extrapolation is mainly useful when the solver operates in PISO mode. Then, the solver starts the SIMPLE loop, iterating until a tolerance criterion is satisfied for both pressure and velocity. The convergence to a given tolerance is enhanced trough the velocity under-relaxation limiting oscillation between correctors. \textcolor{black}{In standard OpenFOAM solver \textit{pimpleFoam}, the pressure relaxation is operated before the velocity update. This procedure can excessively slow down the numerical convergence since the velocity is already relaxed during the update (see Equations \ref{eq:momentumBDF} or \ref{eq:momentumRK}). Hence, the pressure relaxation should be moved after the velocity update. By doing so, the pressure relaxation affects the momentum predictor of the next iteration and the non-orthogonal correction for the Poisson equation}. For each SIMPLE corrector, the PISO loop is executed for a given number of iteration (usually between 4 to 8). After the PISO loop, turbulence equations are solved and, finally, cell and face residuals are calculated for the next Runge-Kutta stage. Note that before resolving turbulence equations, the time step is modified to be consistent with the current Runge-Kutta time step stage.
 
 The segregated sequence of \textit{incompressibleFoam} is detailed in the algorithm below.

	\begin{lstlisting}[caption={\textit{incompressibleFoam} algorithm with PIMPLE approach},label={lst:incompressibleFoamAlgo},language=C++]
	while %$t < t_{end}$) do:
	
	  do Runge-Kutta %iteration):
        %Option: Extrapolate fields (equation \ref{eq:fieldExtrapolation}))
      
	      while %$\epsilon_p < tol_p$ and $\epsilon_\textbf{U} < tol_\textbf{U}$ (SIMPLE correctors)):
	         %Build \textbf{U} equation (\ref{eq:momentumBDF} or \ref{eq:momentumRK}))
	         %Option: relax and solve \textbf{U} equation)
	         	       
	         do PISO %correctors):
              %Update \textbf{H} operator)
              %Momentum interpolation (equations \ref{eq:faceMomentumBDF} or \ref{eq:faceMomentumDIRK}))            
              %Solve Pressure Poisson equations (\ref{eq:poissonBDF} or \ref{eq:poissonDIRK}))
              %Update $\phi$ and $\textbf{U}$)
              %Option: relax p)
           %Solve turbulence equations)
           if last SIMPLE %corrector):
              %Update cell and face residual (equations \ref{eq:residuals} and \ref{eq:faceResiduals}))             
    \end{lstlisting} 

%%%%%%%%%%%%%%%%%%%%%%%%%%%%%%%%%%%%%%%%%%%%%%%%%%%%%%%%%%%%%%%%%%%%%%%%%%%%%%%%%%%%%%%%%%%%%%
%%%%%%%%%%%%%%%%%%%%%%%%%%%%%%%%%%%%%%%%%%%%%%%%%%%%%%%%%%%%%%%%%%%%%%%%%%%%%%%%%%%%%%%%%%%%%%
%%%%%%%%%%%%%%%%%%%%%%%%%%%%%%%%%%%%%%%%%%%%%%%%%%%%%%%%%%%%%%%%%%%%%%%%%%%%%%%%%%%%%%%%%%%%%%

\subsection{Implemented temporal schemes}\label{timeSchemes}
The BDF integrations up to order 3 (section \ref{BDF}) are implemented. For the DIRK method, two SDIRK and ESDIRK schemes, \cite{Kim2022}, are coded: 

\begin{itemize}
    \item SDIRK22 (stage: 2 ; order 2)
    \item SDIRK33 (stage: 3 ; order 3)
    \item CrankNicolson (ESDIRK12) (stage: 1 ; order 2). \textcolor{black}{The resulting implementation is different to the one used in OpenFOAM. Whereas OpenFOAM calculates the residual using the derivative of the last time step field we evaluate it in a consistent way with expression \ref{eq:residuals}}
    \item ESDIRK23 (stage: 2 ; order 3)
\end{itemize}
The coefficients $a_{ij}$ of each scheme are given in section \ref{Appendix}. For steady-state oriented simulations, the \textit{localEuler} scheme based on Local Time Stepping approach and \textit{steadyState} scheme are implemented, although not \textcolor{black}{discussed} in this work.

%%%%%%%%%%%%%%%%%%%%%%%%%%%%%%%%%%%%%%%%%%%%%%%%%%%%%%%%%%%%%%%%%%%%%%%%%%%%%%%%%%%%%%%%%%%%%%
%%%%%%%%%%%%%%%%%%%%%%%%%%%%%%%%%%%%%%%%%%%%%%%%%%%%%%%%%%%%%%%%%%%%%%%%%%%%%%%%%%%%%%%%%%%%%%
%%%%%%%%%%%%%%%%%%%%%%%%%%%%%%%%%%%%%%%%%%%%%%%%%%%%%%%%%%%%%%%%%%%%%%%%%%%%%%%%%%%%%%%%%%%%%%

\subsection{Source code}
The source code of \textit{incompressibleFoam} is available in the following github repository : https://github.com/ferrop/incompressibleFoam.

%%%%%%%%%%%%%%%%%%%%%%%%%%%%%%%%%%%%%%%%%%%%%%%%%%%%%%%%%%%%%%%%%%%%%%%%%%%%%%%%%%%%%%%%%%%%%%
%%%%%%%%%%%%%%%%%%%%%%%%%%%%%%%%%%%%%%%%%%%%%%%%%%%%%%%%%%%%%%%%%%%%%%%%%%%%%%%%%%%%%%%%%%%%%%
%%%%%%%%%%%%%%%%%%%%%%%%%%%%%%%%%%%%%%%%%%%%%%%%%%%%%%%%%%%%%%%%%%%%%%%%%%%%%%%%%%%%%%%%%%%%%%

\section{Test cases}
The performances of the new solver \textit{incompressibleFoam} are tested for three test cases. This section aims to verify the code correctness by comparing the results with experimental data and to obtain additional results regarding MI, pressure form and temporal schemes. For each test case, sensitivity analysis on numerical assumptions (MI, pressure form, time schemes) and solvers (\textit{incompressibleFoam}/\textit{pimpleFoam}) are carried out. The first case is the Taylor-Green vortex flow to assess the conservative properties of different numerical configurations. The second test case is the laminar 2D cavity flow, where a sensitivity analysis on the Reynolds number, self-convergence tests and a comparison with experimental data are performed. Finally, the flow around a cylinder at $R_e = 100$ is studied and the behavior of drag/lift coefficients is compared to the reference data depending on the time step and the numerical configuration.

%%%%%%%%%%%%%%%%%%%%%%%%%%%%%%%%%%%%%%%%%%%%%%%%%%%%%%%%%%%%%%%%%%%%%%%%%%%%%%%%%%%%%%%%%%%%%%
%%%%%%%%%%%%%%%%%%%%%%%%%%%%%%%%%%%%%%%%%%%%%%%%%%%%%%%%%%%%%%%%%%%%%%%%%%%%%%%%%%%%%%%%%%%%%%
%%%%%%%%%%%%%%%%%%%%%%%%%%%%%%%%%%%%%%%%%%%%%%%%%%%%%%%%%%%%%%%%%%%%%%%%%%%%%%%%%%%%%%%%%%%%%%

\subsection{Taylor-Green vortex flow}\label{TGV}
\begin{figure}[h!]
	\centering
	\includegraphics[scale=0.4]{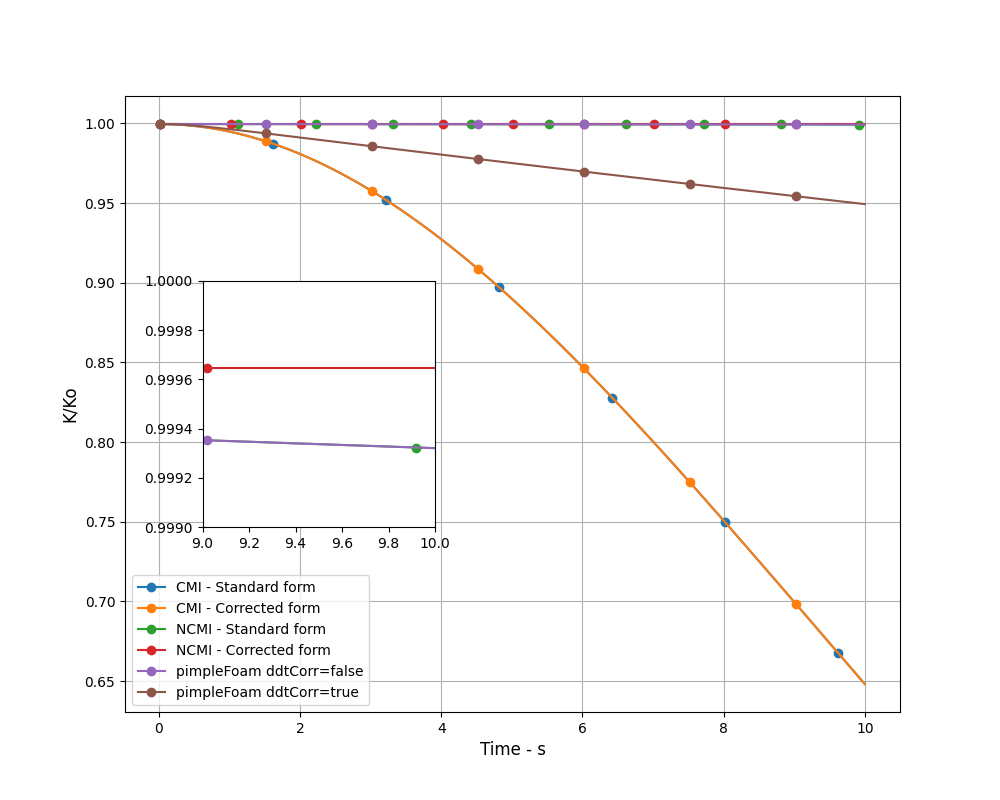}
	\caption{Time evolution of normalized total kinetic energy. Six numerical configurations are compared: CMI/NCMI and \textit{standard}/\textit{corrected} form for Pressure Poisson equation as well as \textit{ddtCorr} (true/false) option for \textit{pimpleFoam} solver.}
	\label{K/Ko}
\end{figure}

The two-dimensional Taylor-Green vortex is a well known test case to assess the conservative properties of a numerical scheme. The inviscid limit is studied here ($\nu = 0$, $R_e= \infty $) where the total kinetic energy $K$ is conserved ($\frac{dK}{dt} = 0$). A sensitivity analysis is performed depending on the numerical configuration. With \textit{incompressibleFoam} solver, four configurations are tested depending on the momentum interpolation and pressure equation forms detailed in sections \ref{momentumInterpolation} and \ref{poisson}. With the \textit{pimpleFoam} solver, the effect of \textit{ddtCorr} option is also evaluated. For each simulation, computed on a 64x64 cartesian grid, the time step is fixed at 0.01 sec ($C_{o_{max}} = 0.05$) and the chosen time integration scheme is the second order \textit{backward} scheme. The convective and Laplacian terms are discretized with a 2nd order centered scheme (\textit{Gauss linear}). Gradients are calculated with the cell centered Gauss method - \textit{Gauss linear} scheme. The pressure velocity coupling is solved with PIMPLE algorithm with the following tolerance error: $\epsilon_p < 10^{-12}$ and $\epsilon_\textbf{U} < 10^{-12}$. The time evolution of the normalized total kinetic energy is shown in Figure \ref{K/Ko}. As observed by \cite{Bartholomew2018}, the results show that the CMI is the most dissipative approach \textcolor{black}{because of the additional diffusive term.
Moreover, the results are almost identical between both pressure forms. This is explained because the face velocity error is $O(d x^2)$ for the CMI irrespective of the pressure form}. The OpenFOAM's MI with \textit{ddtCorr=true} is also quite dissipative but less than CMI. Finally, the NCMI and OpenFOAM's MI with \textit{ddtCorr=false} are the less dissipative, especially the NCMI, with pressure \textit{corrected} form, being almost free of dissipation. This results are in perfect agreement with the one reported by \cite{Komen2020}. 

%%%%%%%%%%%%%%%%%%%%%%%%%%%%%%%%%%%%%%%%%%%%%%%%%%%%%%%%%%%%%%%%%%%%%%%%%%%%%%%%%%%%%%%%%%%%%%
%%%%%%%%%%%%%%%%%%%%%%%%%%%%%%%%%%%%%%%%%%%%%%%%%%%%%%%%%%%%%%%%%%%%%%%%%%%%%%%%%%%%%%%%%%%%%%
%%%%%%%%%%%%%%%%%%%%%%%%%%%%%%%%%%%%%%%%%%%%%%%%%%%%%%%%%%%%%%%%%%%%%%%%%%%%%%%%%%%%%%%%%%%%%%

\subsection{Laminar lid-driven cavity flow}\label{cavityChapter}
The second test case is the laminar lid-driven cavity flow at $R_e = 10, 100, 1000$ and $5000$, computed on a 50x50 cartesian grid. The velocity magnitude $U$ at the top of the $L=1$ m size square is 1 m/s. Sensitivity to Reynolds number, missing in previous studies, is performed by sweeping the kinematic viscosity $\nu = \frac{R_e}{LU}$. Changing the viscosity is more suitable than velocity as the Courant number remains identical for all the Reynolds values. The convective and Laplacian terms are discretized with the 2nd order centered scheme (\textit{Gauss linear}). Gradients are calculated with the cell centered Gauss method - \textit{Gauss linear} scheme. The pressure velocity coupling is solved with PIMPLE and PISO algorithms. In the case of the PIMPLE approach, for each time step or Runge-Kutta stage, the algorithm iterates until $\epsilon_p < 10^{-12}$ and $\epsilon_\textbf{U} < 10^{-12}$. The low tolerance value is necessary to eliminate iterative errors. In the case of the PISO loop, the 2nd order field extrapolation (equation \ref{eq:fieldExtrapolation}) is used. In all the cases, the number of PISO correctors is set to 8. The same configurations used in section \ref{TGV} are tested. The temporal consistency is assessed by performing a self-convergence study for the velocity field magnitude using the L2 norm for the seven time integration schemes listed in section \ref{timeSchemes}. The L2 norm is calculated at $t = 0.1$ sec. For each scheme, seven time steps are calculated and the exact solution is obtained for the smallest one ($10^{-4}$ sec - Courant number  $\approx 10^{-4}$).

\begin{figure}[h!]
	\centering
	\includegraphics[scale=0.7]{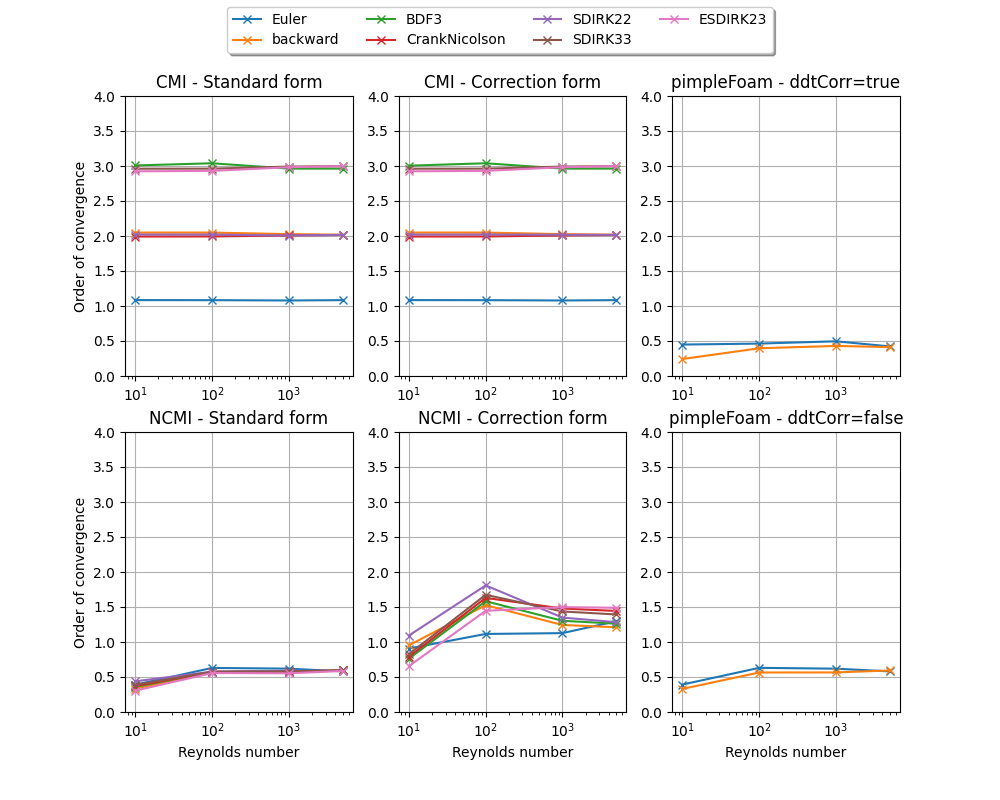}
	\caption{Order of convergence of time integration schemes with \textbf{PIMPLE} algorithm. Six numerical configurations are compared: CMI or NCMI and \textit{standard} or \textit{corrected} form for Pressure Poisson equation as well as \textit{pimpleFoam} with \textit{ddtCorr} option.}
	\label{cavityPimple}
\end{figure}

\begin{figure}[h!]
	\centering
	\includegraphics[scale=0.7]{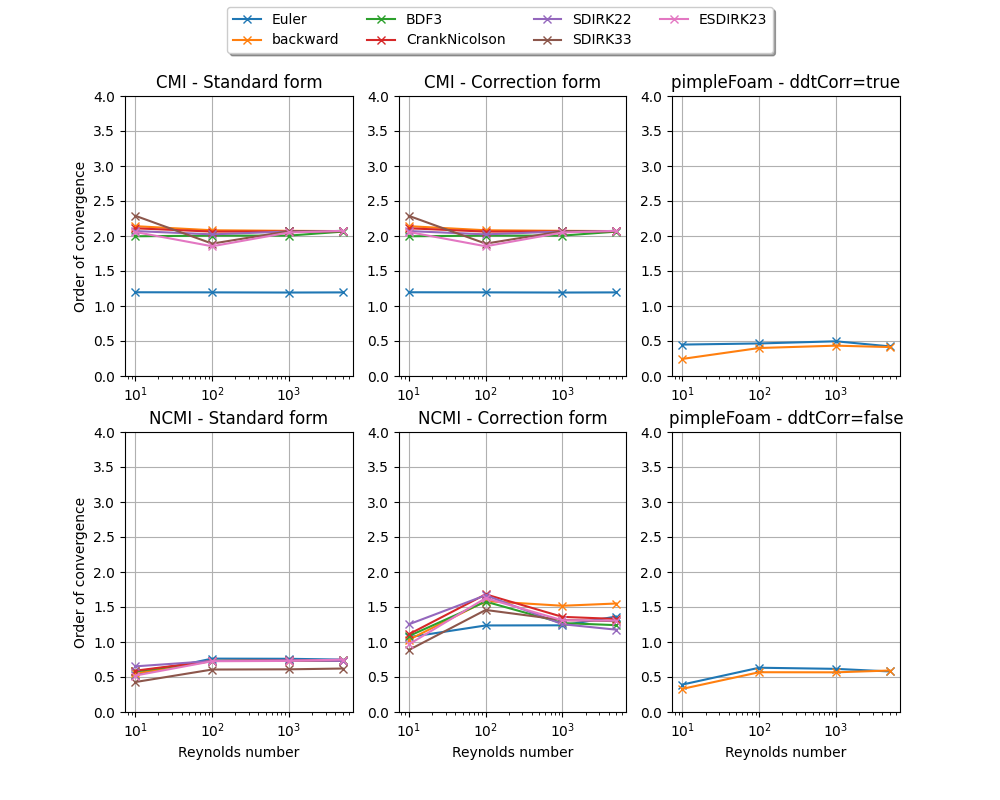}
	\caption{Order of convergence of time integration schemes with \textbf{PISO} algorithm. Six numerical configurations are compared: CMI or NCMI and \textit{standard} or \textit{corrected} form for Pressure Poisson equation as well as \textit{pimpleFoam} with \textit{ddtCorr} option.}
	\label{cavityPISO}
\end{figure}

\begin{itemize}
	\item \textbf{PIMPLE} approach: Figure \ref{cavityPimple} shows the convergence order for the six configurations detailed above. Only the CMI preserves the theoretical accuracy order of all the temporal schemes. No particular difference is observed with either the \textit{standard} or \textit{corrected} form. \textcolor{black}{Both observations are explained by the additional error in $O(d x^2)$ dominating the error in $O(d t^kd x^2), k =1,2$}. With the CMI, the convergence order is independent of the Reynolds number. In contrast, with NCMI, all the schemes fall below the 1st order of convergence with the \textit{standard} Poisson form whereas with the \textit{corrected} form, the order \textcolor{black}{of} convergence is approximately between 1 and 1.5. This results are in line with the previous studies of \cite{KazemiKamyab2015} and \cite{Komen2021}. However, \cite{Komen2021} were able to obtain 2nd order convergence with the pressure \textit{corrected} form for the same case but with a self-convergence study on total kinetic energy. With \textit{pimpleFoam} solver, the results showed a reduction of the convergence order as expected since the OpenFOAM interpolation is non-consistent.
	\item \textbf{PISO} approach: Figure \ref{cavityPISO} highlights that, with CMI, 2nd order convergence is achieved for 2nd order schemes (\textit{backward}, \textit{Crank-Nicolson} and \textit{DIRK22}). For 3rd order schemes (\textit{BDF3}, \textit{DIRK33} and \textit{EDIRK23}), the convergence falls to 2nd order (as expected since the fields extrapolation is only 2nd order). A well-known result is that without the fields extrapolation, the PISO approach leads to first order time accurate results \cite{Issa1986}. For the NCMI and OpenFOAM, the same behavior than with PIMPLE loop is observed. 
\end{itemize}


\begin{figure}[h!]
	\centering
	\includegraphics[scale=0.4]{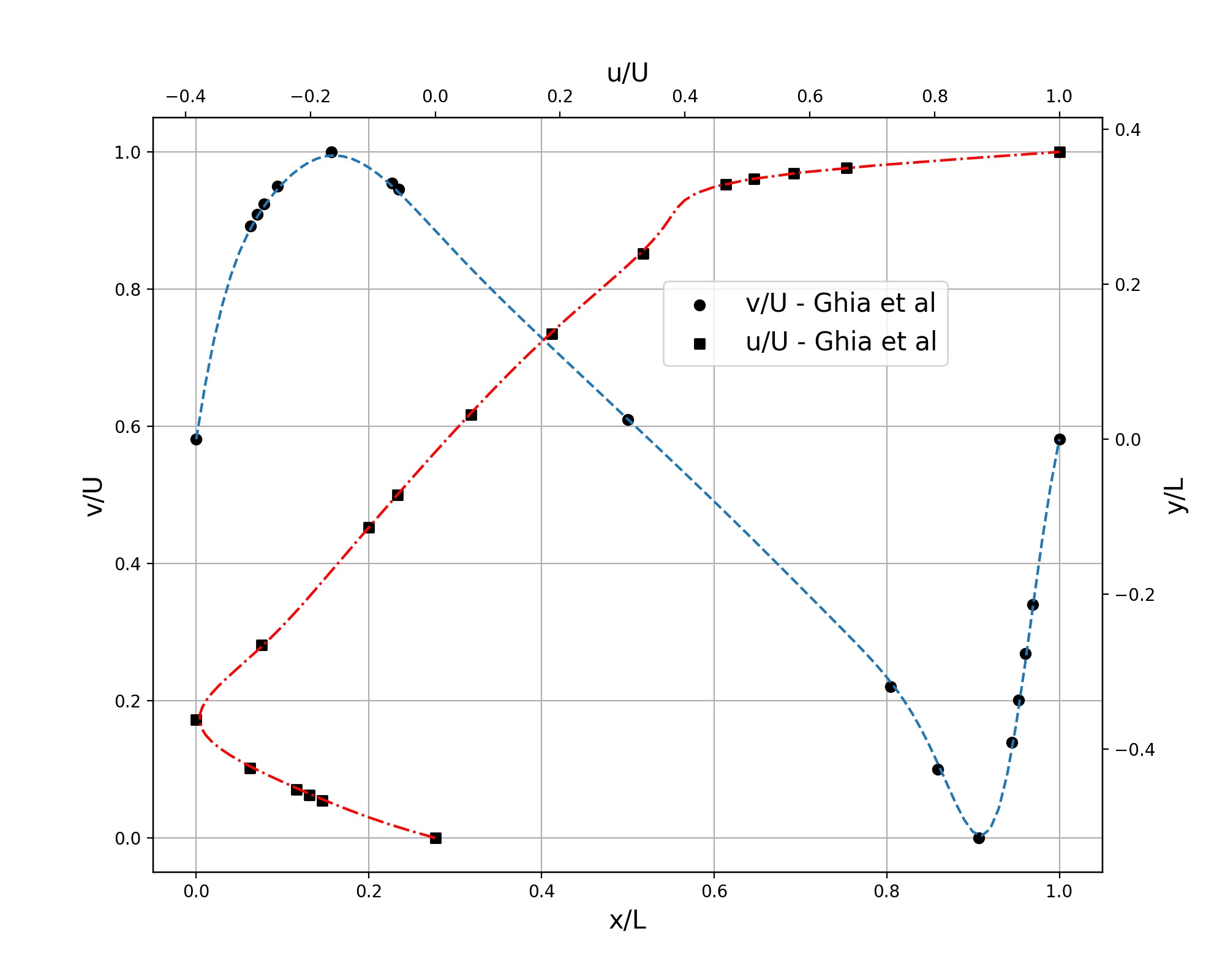}
	\caption{Comparison between the reference solution of \cite{Ghia1982} and \textit{incompressibleFoam} with BDF2 (\textit{backward}) schemes, CMI and pressure \textit{standard} form at t = 50 sec. The red curve is the vertical velocity profile crossing center cavity. The blue curve is the horizontal velocity profile crossing center cavity.}
	\label{cavityResults}
\end{figure}

\begin{figure}[h!]
	\centering
	\includegraphics[scale=0.2]{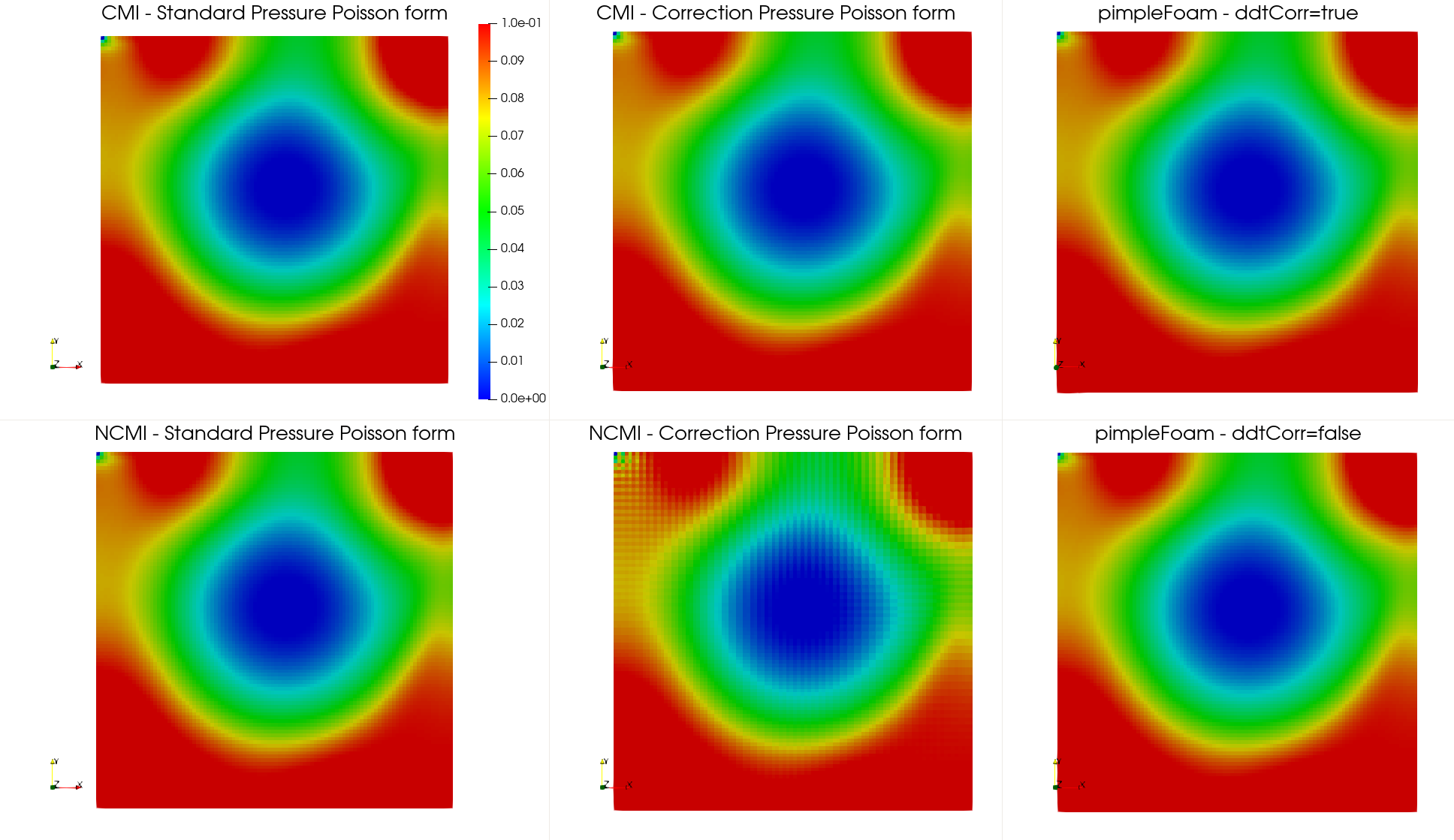}
	\caption{Comparison between the kinematic pressure fields (t = 10 sec) for the six tested configurations. Results are presented for the \textit{backward} scheme. Pressure Checker-boarding occurs for NCMI (pressure \textit{corrected} form).}
	\label{cavityResultsPressure}
\end{figure}

For $R_e = 1000$, numerical simulations have been performed up to steady-state regime (t = 50 sec) with increased mesh resolution (100x100). The results are compared to the reference solution of \cite{Ghia1982} and presented in Figure \ref{cavityResults}. No significant difference was observed regarding the time scheme and/or the numerical configuration, as well as the solver (\textit{pimpleFoam}/\textit{incompressibleFoam}). Hence, only the result obtained with \textit{incompressibleFoam} and BDF2 (\textit{backward}) scheme is presented here. Although the agreement with reference solution is excellent, the configuration with NCMI and pressure \textit{corrected} form exhibits pressure checker-boarding as shown in Figure \ref{cavityResultsPressure}. \textcolor{black}{
This observation is due to the square of the time step, meaning the filter has less weight to damp out the pressure oscillations.
% by the face velocity error $O(d t^2 d x^2)$ being insufficient to damp out the pressure oscillations.
} 


\clearpage

%%%%%%%%%%%%%%%%%%%%%%%%%%%%%%%%%%%%%%%%%%%%%%%%%%%%%%%%%%%%%%%%%%%%%%%%%%%%%%%%%%%%%%%%%%%%%%
%%%%%%%%%%%%%%%%%%%%%%%%%%%%%%%%%%%%%%%%%%%%%%%%%%%%%%%%%%%%%%%%%%%%%%%%%%%%%%%%%%%%%%%%%%%%%%
%%%%%%%%%%%%%%%%%%%%%%%%%%%%%%%%%%%%%%%%%%%%%%%%%%%%%%%%%%%%%%%%%%%%%%%%%%%%%%%%%%%%%%%%%%%%%%

\subsection{Flow around cylinder at $R_e$ = 100}
\begin{figure}[h!]
	\centering
	\includegraphics[scale=0.15]{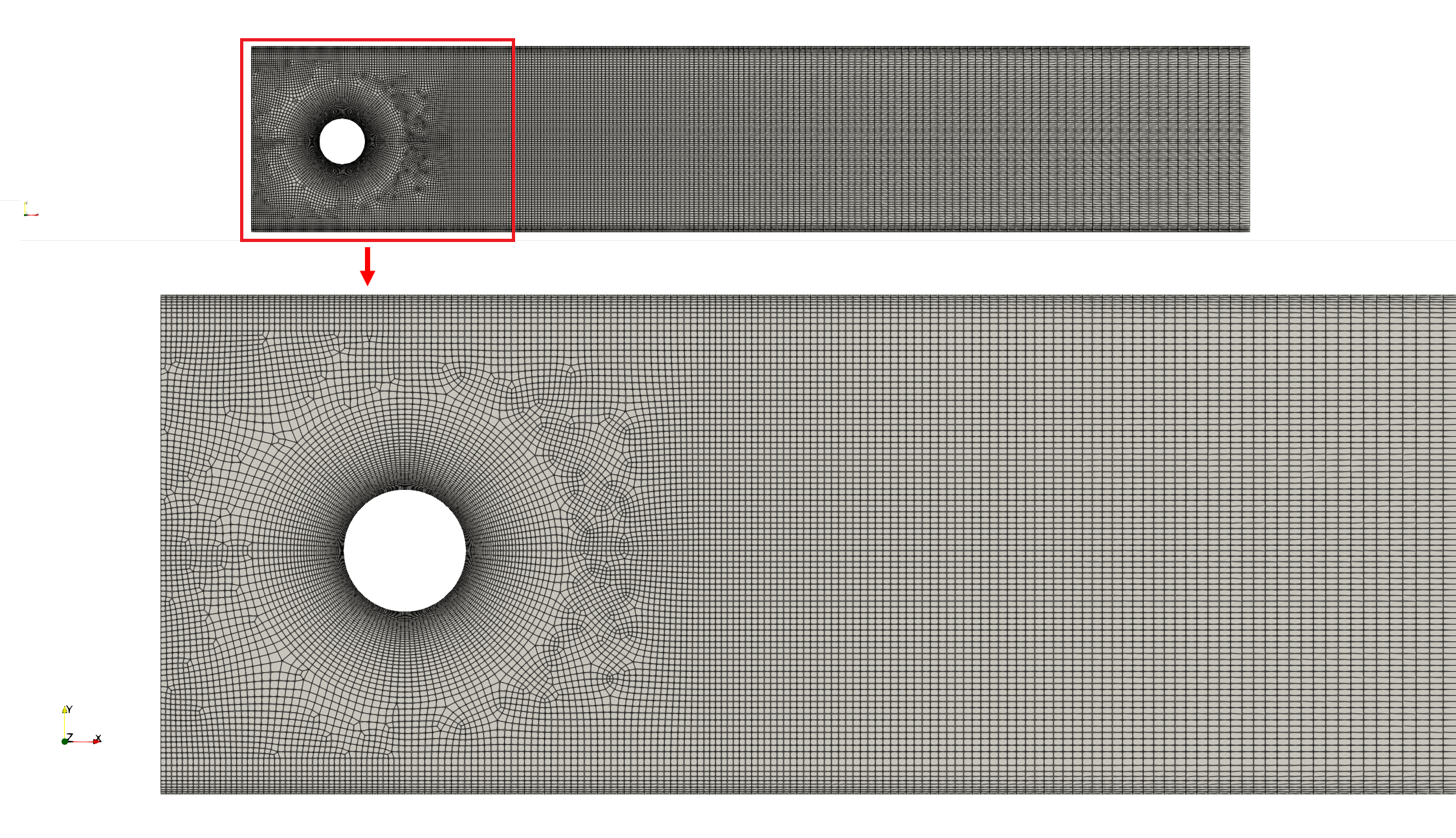}
	\caption{Hybrid grid for laminar flow around cylinder at $R_e$ = 100. The mesh is composed of 24k control volumes.}
	\label{meshCylinder}
\end{figure}

This section is devoted to the 2D-2 test case of \cite{Schaefer1996} where a flow around a cylinder at $R_e = \frac{DU_0}{\nu} = 100$ is considered. At this Reynolds, vortex shedding occurs behind the cylinder. A parabolic velocity profile is imposed at the inlet with a maximum value of $U_{max} = 1.5$ m/s, resulting in an average velocity of $U_0 = 1.0$ m/s. The fluid kinematic viscosity $\nu$ is 0.001 m²/s. 
A hybrid grid \textcolor{black}{similar to the one used in \cite{tukovic2018consistent}'s study}, shown in Figure \ref{meshCylinder}, is built with 24 k control volumes (maximal non-orthogonality of 22 ° and maximal skewness of 0.6).

The convective term is discretized with a 2nd order scheme that combines the \textit{linearUpwind} and the \textit{linear} schemes with a weight of 0.5. The Laplacian term is discretized with the 2nd order centered scheme (\textit{Gauss linear}) with non-orthogonal correction. The gradients are calculated with a least squares (\textit{leastSquares}) scheme. The pressure velocity coupling is solved with the PIMPLE algorithm using the following convergence criteria: $\epsilon_p < 10^{-10}$ and $\epsilon_\textbf{U} < 10^{-10}$. With \textit{incompressibleFoam}, simulations are conducted with four numerical configurations (the same as in section \ref{cavityChapter}), seven time integration schemes (see section \ref{timeSchemes}) and three time steps ($d t = 10^{-4}, 10^{-3}$ and $10^{-2}$ sec, respectively corresponding to a maximal Courant number of $0.1, 1$ and $10$). A relaxation factor of 0.7 is applied for the velocity, stabilizing the solution for the largest time step. With \textit{pimpleFoam}, the simulations are computed with \textit{Euler}/\textit{backward} schemes and with or without \textit{ddtCorr} option. 

\begin{figure}[h!]
	\centering
	\includegraphics[scale=0.5]{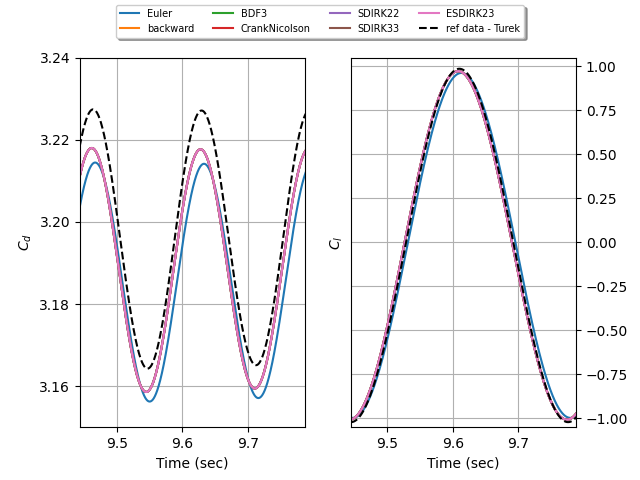}
	\caption{Drag (left figure) and lift (right figure) coefficients history for the smallest time step $d t = 10^{-4}$ sec and comparison to reference data. Numerical configuration : CMI and \textit{standard} pressure form. }
	\label{dragAndLiftTimeStep}
\end{figure}

\begin{figure}[h!]
	\centering
	\includegraphics[scale=0.5]{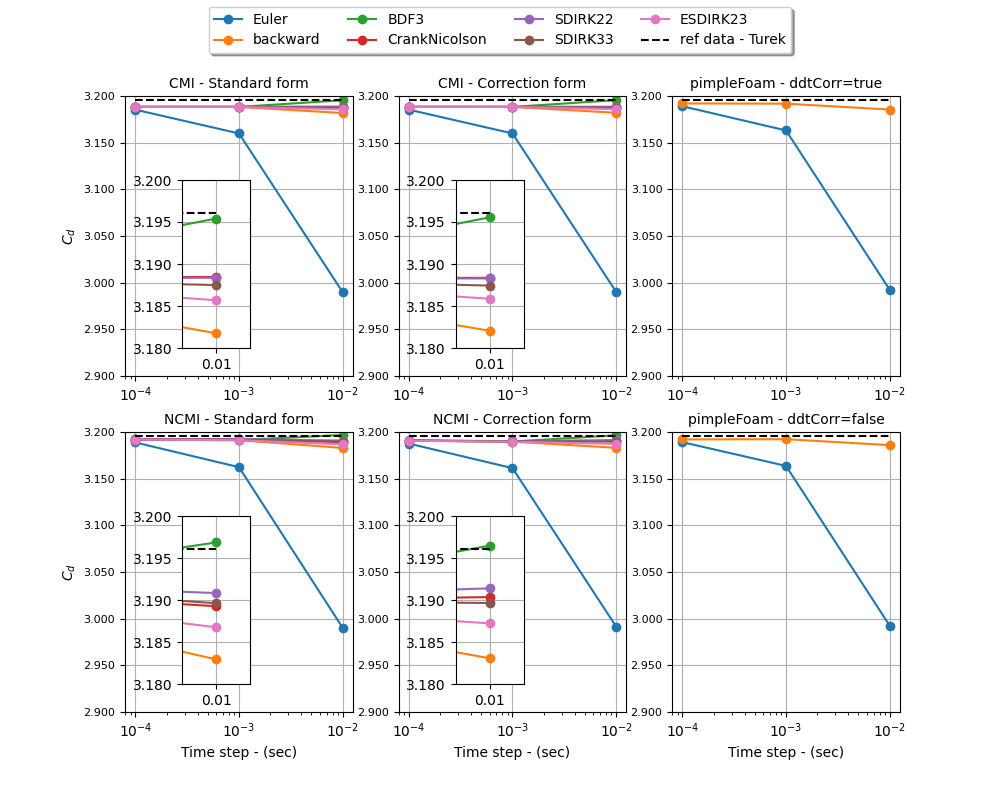}
	\caption{Average drag coefficient as function of time step. Six numerical configurations are compared: CMI/NCMI and \textit{standard}/\textit{corrected} form for Pressure Poisson equation as well as \textit{ddtCorr} (true/false) option for \textit{pimpleFoam} solver.}
	\label{CdAsTimeStep}
\end{figure}

\begin{figure}[h!]
	\centering
	\includegraphics[scale=0.5]{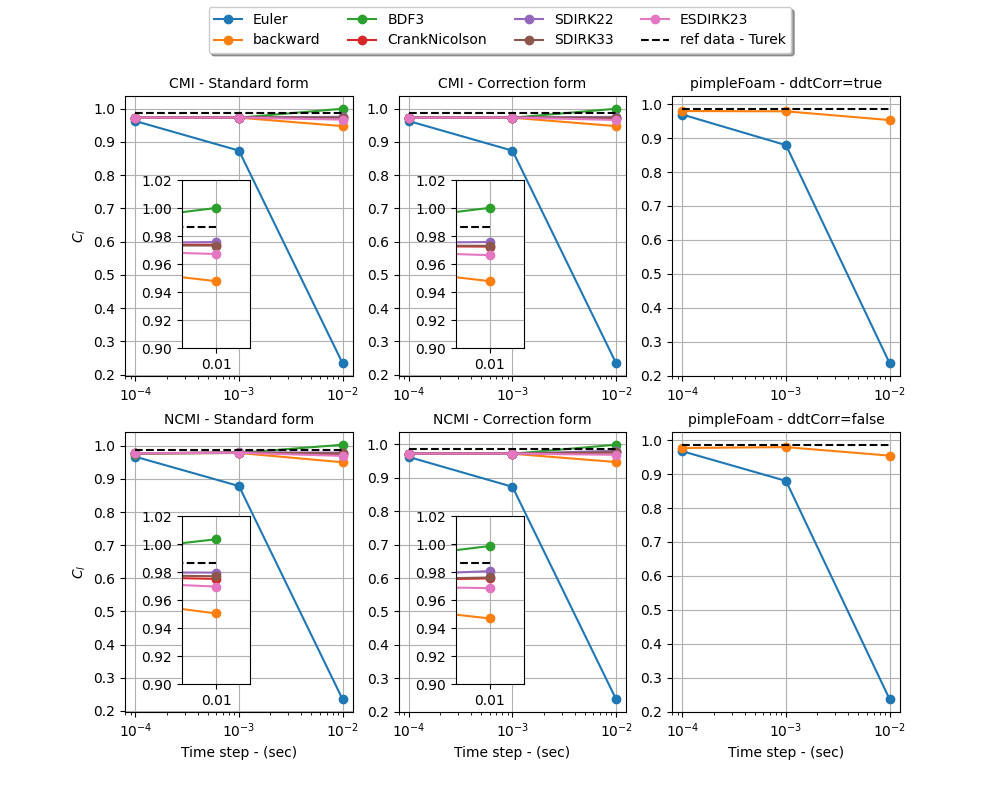}
	\caption{Maximal lift coefficient as function of time step. Six numerical configurations are compared: CMI/NCMI and \textit{standard}/\textit{corrected} form for Pressure Poisson equation as well as \textit{ddtCorr} (true/false) option for \textit{pimpleFoam} solver.}
	\label{maxClAsTimeStep}
\end{figure}

\begin{figure}[h!]
	\centering
	\includegraphics[scale=0.2]{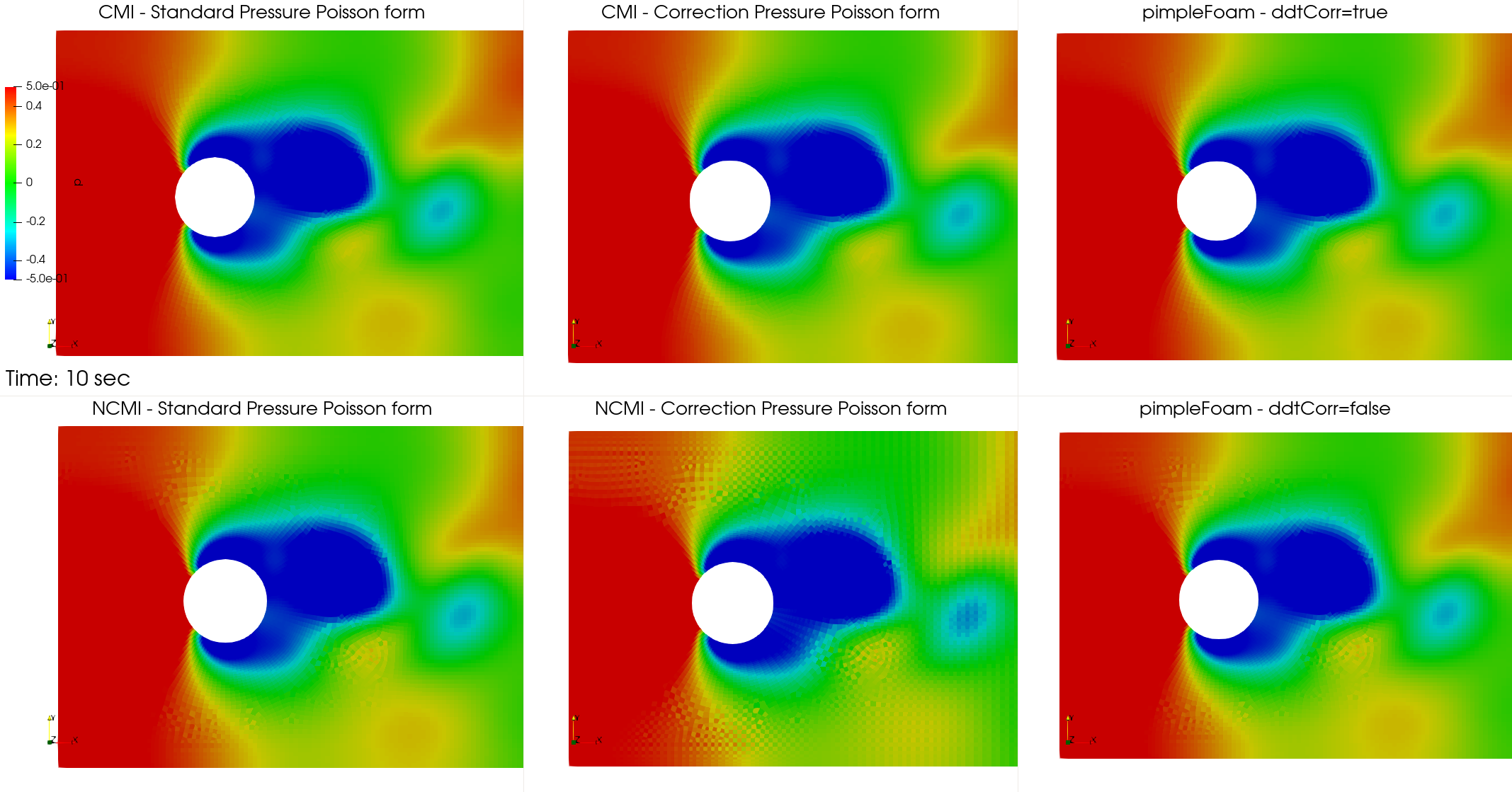}
	\caption{Comparison between the kinematic pressure fields (t = 10 sec) for the six tested configurations. Results are presented for the \textit{backward} scheme. Pressure Checker-boarding occurs for NCMI (both form of Poisson equation) and for \textit{pimpleFoam} with ddtCorr=false.}
	\label{cylinderResultsPressure}
\end{figure}

Since all the results are very close, drag and lift coefficient evolutions are shown in Figure \ref{dragAndLiftTimeStep} only for the CMI and \textit{standard} pressure form. Although the average drag coefficient is slightly underestimated, the results match quite well with the reference data from \cite{Turek} taken for the finest grid and time step (grid level 6 : 133 k elements ; time step $d t = 0.625$ msec). The average drag and maximum lift coefficients are shown in Figures \ref{CdAsTimeStep} and \ref{maxClAsTimeStep} for each simulation. The benefits of 2nd and higher order schemes over the first order \textit{Euler} scheme are clear. For instance, regardeless of the numerical configuration, the \textit{backward} scheme gives similar results with $d t = 10^{-2}$ than the first order \textit{Euler} scheme with $d t = 10^{-4}$. The (E)SDIRK schemes give results close to the reference data for the largest time step and without stability issue. For the largest time step, BDF2 and BDF3 slightly under-perform in comparison to (E)SDIRK based schemes. Since the Crank-Nicolson (ESDIRK12) scheme is a one Runge-Kutta iteration scheme, it provides the best ratio between accuracy and CPU time. As expected with the CMI approach, the results become time step insensitive as the time step is refined. With NCMI and \textit{pimpleFoam} solver, a time step dependency is observed although it is not significant for this test case. The pressure fields in Figure \ref{cylinderResultsPressure} exhibit typical checker-boarding oscillations for NCMI (with stronger effect for the pressure \textit{corrected} form) and \textit{pimpleFoam} with \textit{ddtCorr=false}. \textcolor{black}{
% The stronger pronouncement of checkerboard with NCMI and pressure \textit{corrected} form is typicaly linked to the error on face velocity being $O(d t^2d x^2)$
The stronger pronouncement of checkerboard with NCMI and pressure \textit{corrected} form is typically caused by the face velocity error being $O(dt^2dx^2)$}. In section \ref{cavityChapter}, it has been shown that either NCMI or OpenFOAM's MI leads to time convergence order below first order. However, in terms of drag coefficients, there is no loss of accuracy. For instance, with the \textit{pimpleFoam} solver, the \textit{backward} scheme clearly produces results that are more accurate than the Euler \textit{scheme} and that are very close to those obtained with CMI, which preserves the temporal convergence order.


\clearpage

%%%%%%%%%%%%%%%%%%%%%%%%%%%%%%%%%%%%%%%%%%%%%%%%%%%%%%%%%%%%%%%%%%%%%%%%%%%%%%%%%%%%%%%%%%%%%%
%%%%%%%%%%%%%%%%%%%%%%%%%%%%%%%%%%%%%%%%%%%%%%%%%%%%%%%%%%%%%%%%%%%%%%%%%%%%%%%%%%%%%%%%%%%%%%
%%%%%%%%%%%%%%%%%%%%%%%%%%%%%%%%%%%%%%%%%%%%%%%%%%%%%%%%%%%%%%%%%%%%%%%%%%%%%%%%%%%%%%%%%%%%%%

\section{Conclusions}
This article details the implementation of a new incompressible solver (\textit{incompressibleFoam}) within an OpenFOAM code \cite{Weller1998}. The new solver, with shared sources, gathers several new numerical methods within a same framework. Two forms of the momentum interpolation (consistent \cite{BoYu2002} and non-consistent) as well as a different pressure formulation (\textit{standard} \cite{chorin1968} and \textit{corrected} \cite{Kan1986}) are implemented. Seven time schemes, from \textcolor{black}{steady-state to} BDF and (E)SDIRK up to third order, have been coded. The different numerical methods have been compared using test cases and comparisons to known results and reference data. The solver results are summarized below:
\begin{itemize}
	\item It has been verified, as expected from \cite{KazemiKamyab2015}, \cite{tukovic2018consistent}, that the CMI allows to preserve the theoretical time scheme convergence orders. \textcolor{black}{Regarding the asymptotic behavior, the face velocity error is time step independent. Hence, no temporal error is added during the momentum interpolation preserving the time scheme convergence orders. }
 	\item In agreement with previous studies \cite{Bartholomew2018},\cite{Komen2020}, the CMI and the OpenFOAM MI lead to a high level of numerical dissipation for the \textcolor{black}{inviscid} Taylor-Green vortex flow. \textcolor{black}{It has been shown that this extra dissipation is caused by the face velocity error being $O(dx^2)$. However, for the other test cases presented in this work, it did not lead to degraded results. Further studies should be conducted to determine in which practical engineering applications (if any) this additional dissipation becomes problematic}.
	\item As expected in the case of a transient flow, the \textit{Euler} scheme is highly dissipative and should be avoided to the benefit of a higher order scheme.
 	\item The OpenFOAM MI approach is time step sensitive. The activation of the \textit{ddtCorr} option allows to avoid pressure velocity decoupling. \textcolor{black}{In contrast, the present solver benefits of time step and relaxation factor independent momentum interpolations that should be usefull for practical engineering applications}.  
	\item The 3th order BDF3, SDIRK33 and ESDIRK23 and the 2nd order SDIRK22 schemes did not present significant improvements from the 2nd order backward and CrankNicolson schemes (even for large time steps). Further studies should be conducted to determine in which context they could be useful. 
        \item For self-convergence tests, it has been verified that the 2nd order field extrapolation \textcolor{black}{(not available in pimpleFoam)} with the CMI allows to preserve a 2nd order convergence with the PISO loop. The 2nd order extrapolation has been successfully extended for DIRK schemes.
	\item The \textit{standard} pressure form should be privileged over the \textit{corrected} one, as no practical difference is observed on the results, while the implementation is more complex with the latter.
	\item Despite being almost dissipation free, the combination of NCMI and the pressure \textit{corrected} form always exhibits checker-boarding oscillations. \textcolor{black}{These oscillations can be explained by the stronger decoupling between pressure and velocity caused by the attenuation of the filter due to the square of the time step}. This result tends to mitigate the conclusion obtained by \cite{Komen2021}, who proposed this approach for energy conservative simulations in complex geometries.
        \item Beside the numerical dissipation, where the consequences on practical simulations remain to be identified, the authors recommend the use of a CMI over NCMI/OpenFOAM approach for two main reasons: time step and relaxation factor insensitivity and to avoid pressure velocity decoupling.
        \item \textcolor{black}{Further studies should be conducted to extend the solver capabilities for (E)SDIRK schemes with moving grids in the manner of \cite{Gillebaart2016}.}
\end{itemize}

\section{Appendix}\label{Appendix}
The $a_{ij}$ coefficients are given in the following matrix : 
Crank-Nicolson scheme:
\begin{equation}
a_{ij} = \begin{pmatrix}
0 & 0 \\
0.5 & 0.5 
\end{pmatrix}
\end{equation}

SDIRK22 scheme:
\begin{equation}
a_{ij} = \begin{pmatrix}
\gamma & 0 \\
1-\gamma & \gamma
\end{pmatrix}
\end{equation}

With $$\gamma = 1-\frac{\sqrt 2}{2}$$

SDIRK33 scheme:
\begin{equation}
a_{ij} = \begin{pmatrix}
\gamma & 0 & 0\\
1-\frac{\gamma}{2} & \gamma & 0 \\
-\frac{6\gamma^2-16\gamma+1}{4} & \frac{6\gamma^2-20\gamma+5}{4} & \gamma
\end{pmatrix}
\end{equation}

With $$\gamma = 0.43586652150845899941601945$$

ESDIRK23 scheme:
\begin{equation}
a_{ij} = \begin{pmatrix}
0 & 0 & 0\\
\frac{3}{4} & \frac{3}{4} & 0 \\
\frac{7}{18} & -\frac{4}{18} & \frac{15}{18}
\end{pmatrix}
\end{equation}

%Bibliography
\bibliography{references}

\end{document}